# An Element of Determinism in a Stochastic Flagellar Motor Switch

Li Xie*, Tuba Altindal[†], and Xiao-Lun Wu*

**Author affiliations:**
*Department of Physics and Astronomy
University of Pittsburgh
3941 O'Hara Street
Pittsburgh, PA, USA, 15260

[†]Department of Physics
Simon Fraser University,
8888 University Drive
Burnaby, BC, Canada, V5A 1S6

**Corresponding author:**
Xiao-Lun Wu
Department of Physics
University of Pittsburgh
3941 O'Hara Street
Pittsburgh, PA, USA, 15260
Office: 412-624-0873
Fax: 412-624-9163
Email: xlwu@pitt.edu

**Abstract**

Marine bacterium *Vibrio alginolyticus* uses a single polar flagellum to navigate in an aqueous environment. Similar to *Escherichia coli* cells, the polar flagellar motor has two states; when the motor is counter-clockwise, the cell swims forward and when the motor is clockwise, the cell swims backward. *V. alginolyticus* also incorporates a direction randomization step at the start of the forward swimming interval by flicking its flagellum. To gain an understanding on how the polar flagellar motor switch is regulated, distributions of the forward $\Delta_f$ and backward $\Delta_b$ intervals are investigated herein. We found that the steady-state probability density functions, $P(\Delta_f)$ and $P(\Delta_b)$, of freely swimming bacteria are strongly peaked at a finite time, suggesting that the motor switch is not Poissonian. The short-time inhibition is sufficiently strong and long lasting, i.e., several hundred milliseconds for both intervals, which is readily observed and characterized. Treating motor reversal dynamics as a first-passage problem, which results from conformation fluctuations of the motor switch, we calculated $P(\Delta_f)$ and $P(\Delta_b)$ and found good agreement with the measurements.

# Significance statement

The polar flagellar motor of marine bacterium *Vibrio alginolyticus* alternates between counter-clockwise (CCW) and clockwise (CW) rotations stochastically. It enables the bacterium to perform chemotaxis when it swims forward (CCW rotation) as well as when it swims backward (CW rotation). We found that right after a motor reversal, there exists a refractory period of $\sim 0.2$ second in which another motor reversal is strongly inhibited. This behavior is significantly different from what is known about *Escherichia coli*'s flagellar motor but can be mimicked by a non-equilibrium thermodynamic model. The significance of the refractory period for bacterial chemotaxis in an oceanic environment is also discussed.

# Introduction

The flagellar motor switch controlled by regulatory proteins is fundamental to bacterial chemotaxis and has broad implications for how large protein complexes work. In *Escherichia coli*, FliG, FliM, and FliN proteins form the C-ring, which is the cytoplasmic end of the rotor. About 26 FliG proteins form the upper part of a C-ring that is essential for torque generation. About 34 FliM-FliN subunits form the lower part of a C-ring that acts as a switch by interacting with CheY-P to control the direction of the motor [1]. When all subunits in the switch assume one conformation, which may be assigned as the active state, the motor rotates in clockwise (CW) direction, and when all subunits in the switch assume the other conformation, assigned as the inactive state, the motor rotates in counter-clockwise (CCW) direction. Remarkably, subunits of such a large protein complex can switch coherently and rapidly between the two conformations. The transition rate from one state, say CW to CCW or vice versa, is regulated by the response regulator CheY-P concentration, [YP], inside the bacterium [2, 3]. Observations of wild-type *E. coli* have supported the view that the flagellar motor switches stochastically in a Poissonian fashion. The Poisson behavior manifests itself in the dwell time ($\Delta_{CW}$ or $\Delta_{CCW}$) probability density functions (PDF) being exponential $P(\Delta_s) = \exp(-\Delta_s/\tau_s)/\tau_s$ with the mean time $\tau_s$, where $s \in \{CW, CCW\}$ [4, 3, 5, 6]. Detailed biochemistry information about interacting proteins in *E. coli*'s chemotaxis network shows that [YP] is determined by external chemical signals and the state of adaption in the network [7, 8]. In order to account for the experimentally observed high sensitivity in chemosensing and fast response, cooperativity in chemoreceptors and the motor switch complex appears to be necessary [9, 10]. The classical theory taking into account these collective effects has been MWC or KNF models [11, 12]. A more general model describing protein conformation spread using Ising spins has been recently introduced [13]. This latter model allows protein conformation fluctuations to be calculated using statistical mechanics methods and is found to be in good agreement with experiments [14, 5]. When these models operate at equilibrium, i.e., constant temperature with constant transition rates between different states, both $P(\Delta_{CW})$ and $P(\Delta_{CCW})$ are sums of exponential functions and decay monotonically [15].

Herein, we report switching statistics of the polar flagellar motor of the marine bacterium *Vibrio alginolyticus* YM4 (Pof$^+$, Laf$^-$) [16]. In an aqueous environment, *V. alginolyticus* expresses a single polar flagellum that is driven by a two-state motor similar to *E. coli* [17].

When the motor turns in the CCW direction, the cell body is pushed by the flagellum, which we called forward (f) swimming, and when the motor turns in the CW direction, the cell body is pulled by the flagellum, which we called backward (b) swimming. However, unlike *E. coli* whose lateral flagellum is connected to its motor by a bent hook, the polar flagellar hook of *V. alginolyticus* is straight but bendable when a thrust is above a certain threshold [18, 19]. Consequently, at the *beginning* of each forward interval, the elastic instability of the flagellar hook induces a bent that can change the cell's movement direction on average by $\Delta\theta \simeq 90^o$. This conspicuous movement was termed a flick [20]. The motility pattern of *V. alginolyticus* is thus a cyclic three-step (forward-backward-flick) process; motor reversals from CCW to CW result in a kink with $\Delta\theta \simeq 180^o$, but reversals from CW to CCW result in a broad range of angles $0 \leq \Delta\theta \leq 180^o$. Moreover, for swimming at low Reynolds numbers, the translational motion of the cell body responds to a change in the thrust force almost instantaneously [21], and the determination of a reversal event is only limited by the temporal resolution of the experimental setup. A bacterial swimming trajectory punctuated by these sequential sharp features permits us to reliably construct a time series of the flagellar motor states [19, 20].

Besides studying motor switching behaviors of free-swimming cells using video microscopy, measurements were also conducted by confining individual bacteria in an optical trap that records a cell's position in the trap at a much higher sampling rate. Despite very different temporal resolutions of the two methods, they yield consistent results showing that the forward $\Delta_f$ and backward $\Delta_b$ dwell-time PDFs, $P(\Delta_f)$ and $P(\Delta_b)$, are strongly peaked at $\sim 270$ and $\sim 370$ ms, respectively. These results together suggest that the polar flagellar motor of *V. alginolyticus* is regulated in a fashion very different from *E. coli*.

## Results

### Statistical Correlations of a Polar Flagellar Motor Switch

The principal finding of our experiment is that *V. alginolyticus*' motor reversal events are mutually exclusive, exhibiting strongly non-Poissonian fluctuations. This behavior suggests that at least one of the steps in the regulation of motor reversal is thermodynamically irreversible [15, 22]. A quick and convenient method to see this unusual behavior is by means of counting statistics commonly employed in the study of quantum particles and inter-spike intervals in neuron dynamics [23, 24, 25]. For the former case, simply counting the particle arrivals at a detector can reveal the quantum nature of the particles. If the particle arrival times are bunched together, they are bosons but if the times are anti-bunched, they are fermions. For the latter case, very useful clues about the underlying neurophysiological processes can be extracted from the observed spike train [25, 26].

Similarly, counting motor switching events could also shed light on the mechanism that regulates the motor direction. A useful quantity characterizing stochastic nature of the flagellar motor switch is the Fano factor, $F = \sigma^2/N$. Here $N$ is the mean count of motor reversals during time $T$, which includes both CCW to CW and CW to CCW reversals, and $\sigma^2$ is the variance. If the transition of the motor between the two states is governed by the equilibrium models such as Ref. [15], $F \geq 1$ for $T \gg \Delta_f, \Delta_b$. To measure $F$, five

*V. alginolyticus* bacteria were randomly picked and each tracked for ∼10 minutes. Each track was then segmented into consecutive intervals of length $T$, and the average number of switches $N(T)$ within $T$, and its variance $\sigma^2(T)$ were calculated. Since counting is an integration process and missing an entire swimming cycle (forward+backward) is statistically unlikely based on our measured dwell-time distributions (to be discussed below), this method is not sensitive to the time resolution of the measurements. Fig. 1(A) displays $\sigma^2$ vs. $N$ for the five cells tracked. It is seen that in all cases $\sigma^2(T)$ is smaller than $N(T)$ or $F < 1$. The observation suggests that two consecutive motor reversal events are mutually exclusive in a fashion akin to fermions and therefore cannot be accounted for by the equilibrium model.

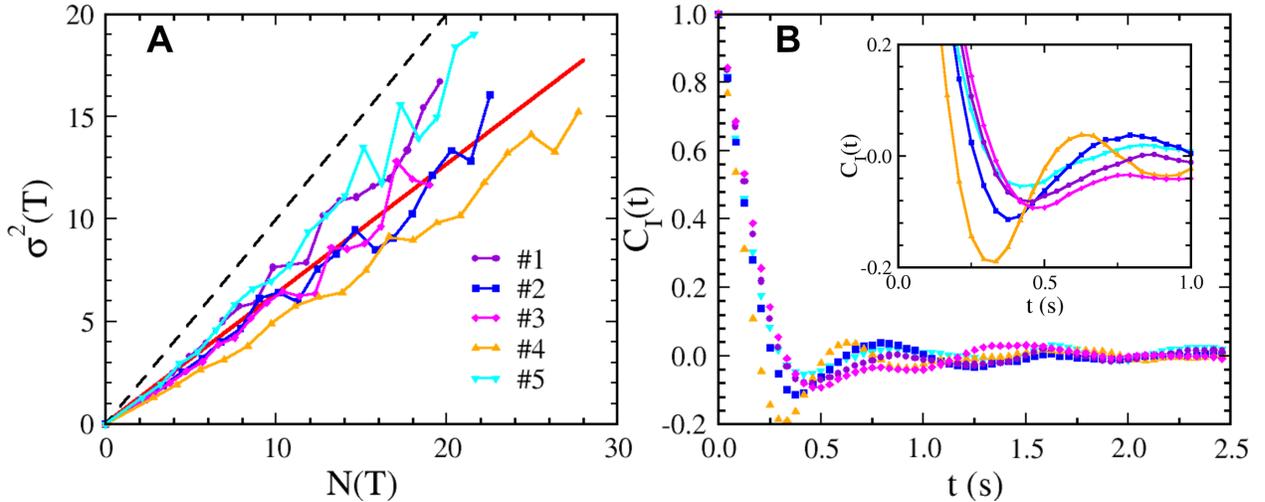

Figure 1: Measurements from five *V. alginolyticus* bacteria each being tracked for ten minutes. (A) $\sigma^2$ vs. $N$ for five bacterial trajectories #1-#5 are displayed as indicated by the legend. The dashed black line represents $\sigma^2/N = 1$. Assuming that $\sigma^2$ vs. $N$ is linear, a linear regression using the data from the five cells yields a straight line with a slope (or Fano factor) of 0.63, which is represented by the red line. (B) The autocorrelation functions $C_I(t)$ computed using the time series $I(t')$ for the five cells (see legend in (A)), where $I(t') = -1$ for CW and $+1$ for CCW motor state (see main text). The inset displays short-time oscillations with more details.

To characterize temporal fluctuations of the observed switching events, two types of correlation functions are computed: In the first, a binary time series $I(t')$ is constructed based on the state of motor rotation with $I = +1$ for CCW and $I = -1$ for CW. The autocorrelation function is defined as,

$$C_I(t) = \frac{\langle I(t')I(t'+t)\rangle - \langle I(t')\rangle^2}{\langle I(t')^2\rangle - \langle I(t')\rangle^2}, \qquad (1)$$

where $\langle ...\rangle$ indicates average over $t'$. If the motor switch is regulated as described by the equilibrium models, $C_I(t)$ decay monotonically with time. However, this is not what was observed in the measurement, which is displayed in Fig. 1(B) for the five time series. We note that although $C_I(t)$ decays with time, it is non-monotonic, showing oscillations with a period slightly less than a second.

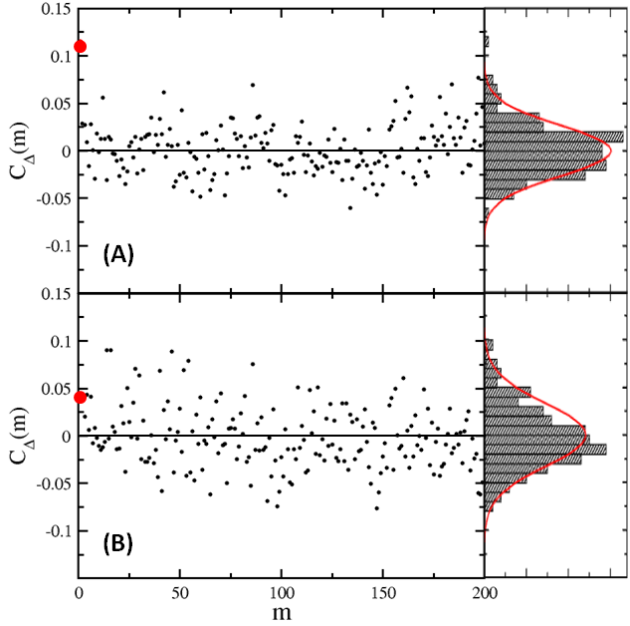

Figure 2: Dwell-time correlations $C_\Delta(m)$ for cells #4 (A) and #5 (B). $C_\Delta(m)$ vs. $m$ are plotted on the left panels and the PDFs of $C_\Delta(m)$ are plotted on the right panels. The big red dots denote $C_\Delta(1)$. The red curves depict normal distributions centered at zero.

The fact that $F < 1$ and $C_I(t)$ oscillates could be a result of temporal correlations in the swimming intervals. This prompts us to examine correlations between dwell times, which is characterized by the second correlation function

$$C_\Delta(m) = \frac{\langle \Delta_i \Delta_{i+m} \rangle - \langle \Delta_i \rangle^2}{\langle \Delta_i^2 \rangle - \langle \Delta_i \rangle^2}, \qquad (2)$$

where $\Delta_i$ is the waiting time between the $ith$ and the $(i+1)th$ switching event, and $\langle ... \rangle$ indicates average over all $i$. By this definition $C_\Delta(0) = 1$, and the next significant correlation is $C_\Delta(1)$, which were found to be 0.11, 0.08, 0.09, 0.11, and 0.04, respectively for cells #1-#5. We also calculated $C_\Delta(m)$ for $m > 1$; the data for the $4th$ and for $5th$ cell, which have the largest and the smallest $F$, are displayed respectively in Fig. 2(A, B). We note that for all fives cells, $C_\Delta(m)$ spread randomly, and their distributions can be mimicked by the normal distributions that center at zero with the standard deviations $\sigma_C$ varying between 0.02 to 0.04 (see Fig. 2). Since $C_\Delta(1)$ is overall greater than $\sigma_C$ but much smaller than $C_\Delta(0) = 1$, the consecutive interval lengths may be correlated, but the correlation is very weak and does not extend beyond $m = 1$. Also, such correlation cannot account for the observed $F < 1$ and oscillations in $C_I(t)$. First, even though the switching sequences of cell #1 and #4 have the same $C_\Delta(1) = 0.11$, we found that $F = 0.77$ for #1 is the largest and $F = 0.53$ for #4 is the smallest among the 5 sequences studied. Second, when random shuffling is applied to the time series, the distributions of $C_\Delta(m)$ are more-or-less unchanged, and the oscillations in $C_I(t)$ remain (see Supporting Information (SI)).

# Dwell-Time Statistics Studied by Video Microscopy

## (A) Dwell-Time PDFs Are Non-monotonic in TMN Motility Buffer

In this set of investigation we focus on the dwell-time PDFs, $P(\Delta_f)$ and $P(\Delta_b)$, of a large ensemble of cells, $n \simeq 500$. Because the measurements depend critically on how precisely individual motor reversal events can be determined, in Materials and Methods we provide detailed information concerning the measurement, the uncertainties, and the expected viscoelastic response times that could smear otherwise sharp transitions between the rotation states. The analysis therein demonstrated that our method can detect the motor reversal moment with adequate precision.

A total of $\sim 500$ cells' trajectories were analyzed resulting in $\sim 700 - 800$ individual $\Delta_f$ and $\Delta_b$. The PDFs $P(\Delta_f)$ and $P(\Delta_b)$ were displayed in Figs. 3(A, B), where the shaded area indicates the lower bound for the interval measurement ($\sim 66$ ms). It is evident that $P(\Delta_f)$ and $P(\Delta_b)$ are strongly peaked at $\Delta_{fmax} \simeq 0.27$ s and $\Delta_{bmax} \simeq 0.37$ s, respectively. The larger $\Delta_{bmax}$ suggests that spontaneous motor reversals in short times are more strongly inhibited in the CW direction than in the CCW direction. Moreover, the broad tails for both $P(\Delta_f)$ and $P(\Delta_b)$ make the distributions skewed towards small intervals. To a very good approximation the tail portion of the PDFs is exponential as is evident by the linear behavior seen on the semi-logarithmic plots of Figs. 3(A, B). By fitting the tails of these distributions using an exponential function, $\exp(-\Delta_s/\tau_s^\infty)$ where $s \in \{f, b\}$, we found $\tau_f^\infty = 0.32$ s and $\tau_b^\infty = 0.27$ s for the forward and backward intervals, respectively.

We note that the measured dwell-time PDFs for *V. alginolyticus* are significantly different from those observed in *E. coli*, which are exponentially distributed [4, 3, 5, 6]. While there is reasonably strong evidence suggesting that *E. coli*'s flagellar motor switching is controlled by thermally activated Poisson processes [27, 3, 5], the flagellar motor switch of the marine bacterium is regulated in a decidedly different fashion. Peaking of the dwell-time PDFs seen in *V. alginolyticus* suggests that the motor switch of *V. alginolyticus* has a refractory period right after the motor has switched, i.e., within this period another motor reversal is strongly inhibited. Swimming interval times in the marine bacterium, therefore, appear to be governed by two competing processes, the short-time inhibition and long-time Poisson-like process. It comes as no surprise therefore that the oscillatory behavior seen in the correlation function $C_I(t)$ is a result of the dominant time scales, $\Delta_{fmax}$ and $\Delta_{bmax}$, in the dwell-time PDFs.

## (B) Dwell-Time Distributions in the Presence of Chemorepellent Are Also Non-monotonic

While the above steady-state measurements are informative, revealing a significantly different switching behavior compared to *E. coli*'s flagellar motor, it is useful to see how $P(\Delta_f)$ and $P(\Delta_b)$ are altered when an external perturbation is applied. Most flagellar motor switches studied so far are controlled by the phosphorylated form of the regulatory protein CheY [28]. For *E. coli*, an elevated [YP] increases the switching probability from the CCW to CW state and hence the CW bias [2, 3, 5]. Regulator CheY in *V. alginolyticus* has a great deal of homology to its *E. coli* counterpart; they are 84% identical [29]. Overexpressing *cheY* or exposure to repellent phenol was shown to make the *V. alginolyticus* cell change swimming directions more frequently [30, 29]. Interestingly, this response to phenol is not adaptive in YM4 since the motor reversal rate changed by less than 10% after a 20-minute exposure [30]. This provides a convenient means to change [YP] in the cells, allowing $P(\Delta_f)$ and $P(\Delta_b)$ to be measured in a new steady state. The measured $P(\Delta_f)$ and $P(\Delta_b)$ from cells in TMN+10 mM phenol are displayed in Figs. 3 (C, D). These distributions are similar to those acquired in TMN (Figs. 3 (A, B)) except here the time scales are significantly shortened with $\Delta_{fmax} = 0.17$ s and $\Delta_{bmax} = 0.27$ s. The exponential tails are characterized by $\tau_f^\infty = 0.25$ s and $\tau_b^\infty = 0.14$ s, respectively.

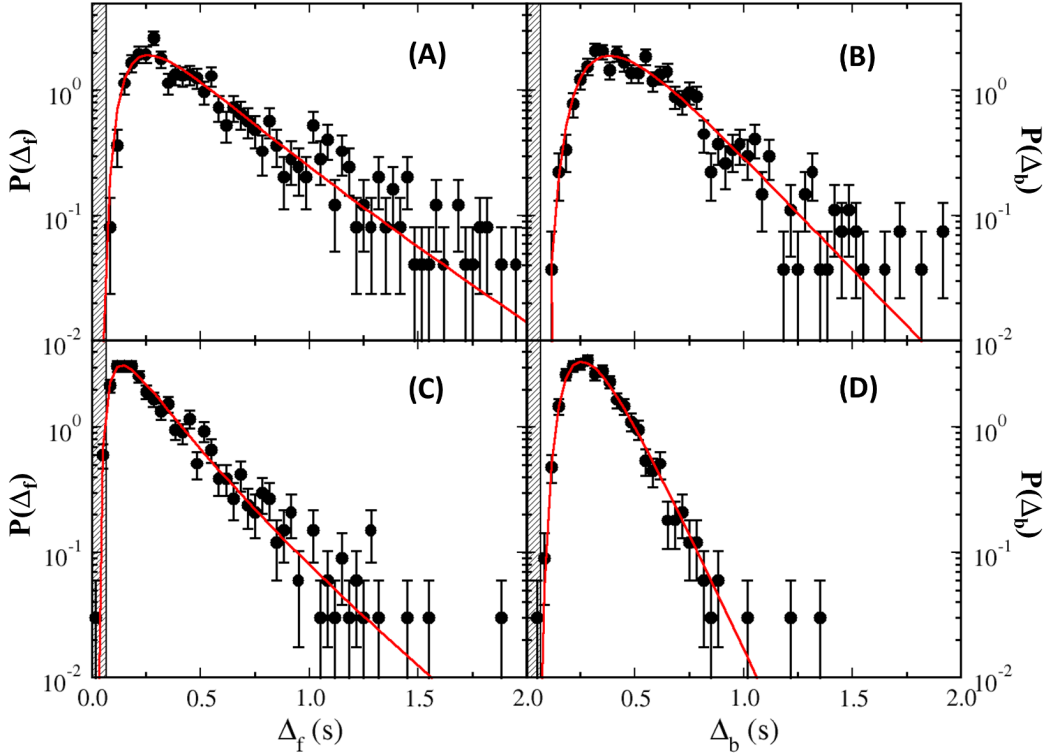

Figure 3: Dwell-time PDFs of YM4 in TMN (A-B) and TMN+phenol (C-D). The measurements and the fitting curves are displayed as black dots and red lines. Specifically, $P(\Delta_f)$ and $P(\Delta_b)$ are fitted by an inverse Gaussian distribution as described in main text. The shaded areas in these figures denote the short-time (66 ms) limitation of our measurements. Note that the time scales for the phenol data are significantly smaller than those measured without phenol.

Despite very different time scales in $P(\Delta_f)$ and $P(\Delta_b)$ measured in the two steady states, it is remarkable that all of them can be described by the inverse Gaussian distribution. This suggests that the underlying regulation mechanisms are identical for the two steady states and for the two intervals. It is also noteworthy that incessant motor reversals at a high rate when *V. alginolyticus* is exposed to phenol is rather peculiar and is at variance with *E. coli*'s response to the same chemical. When a non-adaptive *E. coli* cell is exposed to phenol, its motor is permanently "locked" in the CW direction [31]. This extreme CW bias can be explained as a result of elevated [YP] in the cytoplasm of *E. coli* that "forces" the motor to run exclusively in the CW direction. If exposure to the repellent has the same effect of elevating [YP] in *V. alginolyticus*, for which there is little reason to believe otherwise, an inescapable conclusion is that the flagellar motor switch of *V. alginolyticus* reacts to this regulatory protein very differently from *E. coli* [30, 29, 32]. This important observation motivates a molecular toggle switch model for the polar flagellar motor to be presented in the Theoretical Modeling section.

## Optical Trapping Improves Short-Time Resolution

In this set of measurements we wish to capture fast events that might have escaped detection by video microscopy. This is achieved by capturing individual bacteria in an optical trap, where movements of the bacterium can be sampled at a rate of 10 kHz (more details in SI). In the optical trap, a bacterium has two stable positions, i.e., it can be held either at the tip or at the tail of the cell body depending on its swimming direction (see Figs.4(A-B)). These two positions are readily resolved when the $z$ axis is slightly tilted. In this case, the cell-body position along the $z$ axis has a small projection along the $x$ axis and is recorded by the position-sensitive detector (PSD) as displayed in Fig. 4(C) [33].

To minimize photo-damage, cells were trapped for 3 s, resulting in $5 \sim 10$ switching events per cell. About 320 cells were analyzed with $\sim 3000$ switching events. It is seen that the cell-body position time trace $x(t)$ alternates between two constant levels, $+x_0$ and $-x_0$, with a residence time of $\sim 0.3$ s. The upper and lower states are separated by sharp transitions that can be characterized by a transition time $t_s$. We characterized $t_s$ by measuring the time it takes for the signal to increase from $-0.8x_0$ to $0.8x_0$ or vice versa. The PDF of $t_s$ was constructed using 38 cells with a total of 172 switching events. As shown in Fig. 4(D), $P(t_s)$ peaks at 15 ms and can be adequately fit by a log-normal distribution. The skewness of the distribution makes the mean transition time somewhat larger, $\bar{t}_s \simeq 22$ ms. Considering that the full length of *V. alginolyticus* under our culture condition is 2-3 $\mu$m, and their average swimming speed is $v_{sw} \simeq 55 \, \mu$m/s, the 22 ms transition time suggests that the bacterium moves about half of its body length in the optical trap. The mean transition time sets the temporal resolution of our trapping technique, which is about a factor of three faster than video microscopy. To detect motor reversals from a time trace, $x(t)$ is convoluted with a smooth-differential filter $F(t) = -(t/2c^2)\exp(-t^2/2c^2)$, $\Delta x(t) = \int_{-\infty}^{\infty} F(t-t')x(t')dt'$, and the result is displaed by the red curve in Fig. 4(C), where $c = 30$ ms. The motor reversal moments are then determined whenever a peak or a valley of $\Delta x(t)$ passes the thresholds, which is set at 75% of $\pm x_0$ as indicated by the blue lines in Fig. 4(C). From the sequences of the motor reversal events, the dwell times $\Delta_u$ are calculated, where the subscript "$u$" indicates that the dwell time could be either $\Delta_f$ or $\Delta_b$ because it is not possible to determine the orientation of a cell in the optical trap [33]. Using this method (see more details in SI), PDF of $\Delta_u$ is constructed and presented in Fig. 5.

The measured $P(\Delta_u)$ displays a fast rise for small $\Delta_u$ and an exponential-like tail for large $\Delta_u$; the deviation from a straight line in the tail is due to the fact that the interval measurement contains a mixture of $\Delta_f$ and $\Delta_b$. The measured PDF is consistent with those presented in Fig. 3 (A, B), and the calculated mean dwell time $\sim 0.3$ s also compares well with the data acquired using video microscopy. Because of the higher temporal resolution, we were able to acquire more data for small time intervals. This allowed us to examine how $P(\Delta_u)$ behaves in the limit of small $\Delta_u$. This relationship is significant because it tells us how strongly the short time intervals are inhibited. In the inset of Fig. 5, $P(\Delta_u)$ vs. $\Delta_u$ is plotted on a log-log scale. One observes that for $50 \leq \Delta_u \leq 200\,\text{ms}$, $P(\Delta_u)$ increases quadratically with $\Delta_u$, confirming the non-monotonic behavior seen above.

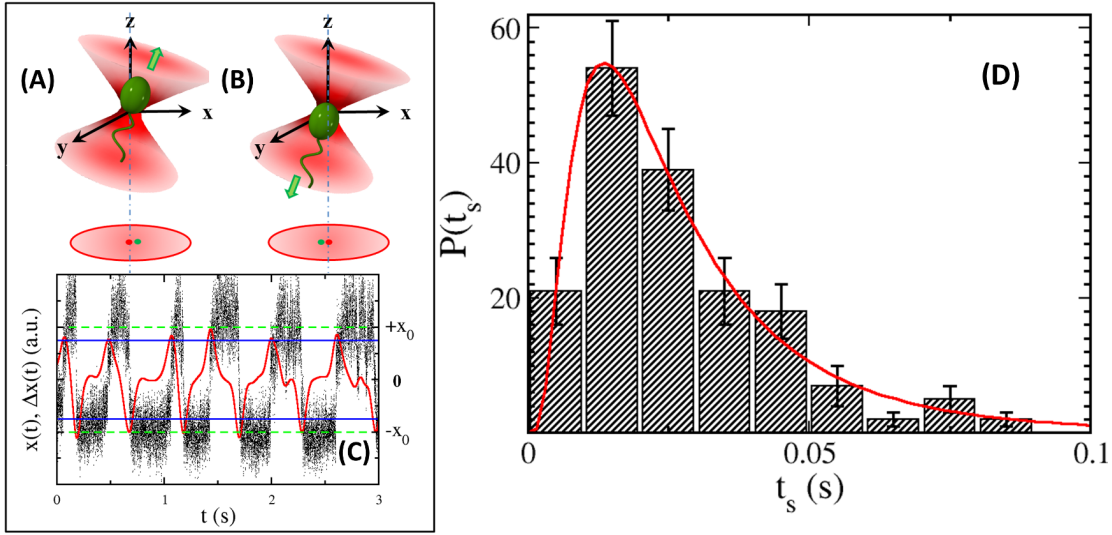

Figure 4: Optical trapping and swimming interval measurements. (A) depicts a forward swimming bacterium trapped at the rear end of the cell body whereas (B) depicts a backward swimming bacterium trapped at the front end of the cell body. In both (A, B), the green arrows indicate the swimming direction of the cell. This results in a shift in the center of mass of the cell in the optical trap, where the dot-dash lines and the red dots indicate the center of the optical trap. By slightly tilting the optical trap, the center of mass location of the cell body has a small projection (the green dots) along the $x$-axis of PSD. As the cell swims back and forth, its $x$ position fluctuates as depicted by the small black dots in (C). There are two stable positions, $+x_0$ and $-x_0$, corresponding to bacterial forward and backward swimming, and they are delineated by the green dashed lines. The red curve represents convoluted time trace $\Delta x(t)$, and the blue solid lines indicate the thresholds. (D) The histogram shows the transition-time distribution $P(t_s)$. The red line is the fitting to the log-normal distribution with the mean $\mu_u \simeq -3.8$ and the standard deviation $\sigma_u \simeq 0.72$.

## Theoretical Modeling

### The Dwell-Time PDFs Are Consistent with a First-Passage Time Distribution

The above experiments establish the following two important facts about the polar flagellar motor of *V. alginolyticus*: (i) Binding of CheY-P to the motor facilitates motor reversal irrespective of its current rotation state. In this sense, it behaves like a toggle switch. (ii) There exists a short refractory period within which a motor reversal is strongly inhibited. These unusual features can be significant for marine bacteria to survive in an oceanic environment and call for their quantitative understanding.

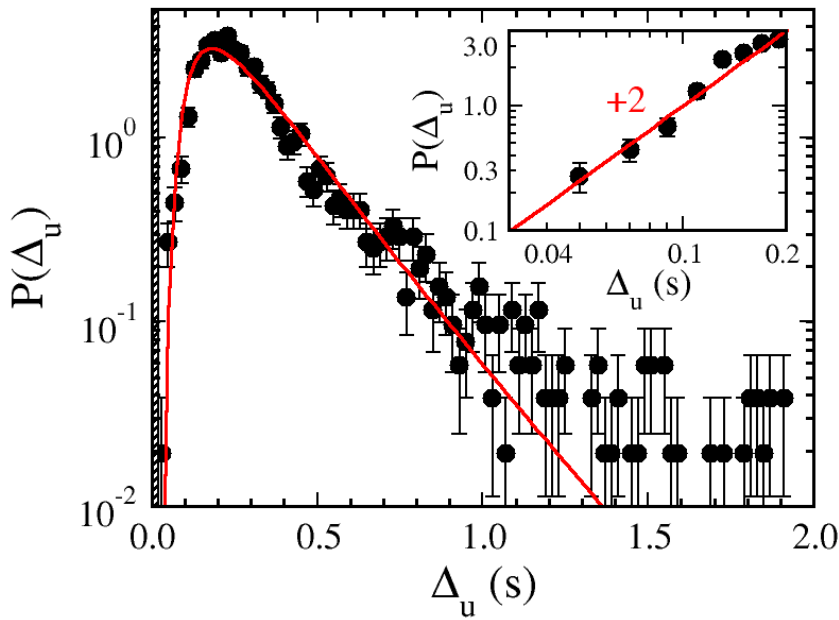

Figure 5: Dwell time-distribution determined by optical trapping. The undifferentiated dwell-time distribution $P(\Delta_u)$ is represented by the black dots. The error bars indicate uncertainties in the measurement and the red line is the fit to $P(\Delta_u)$ using the inverse Gaussian distribution. The shaded area marks the short-time detection limit $\bar{t}_s \simeq 22\,\mathrm{ms}$ discussed in the main text. The inset is a log-log plot of $P(\Delta_u)$ vs. $\Delta_u$ for small intervals. It shows that $P(\Delta_u)$ scales as $\Delta_u^2$.

According to the molecular mechanism proposed by Paul et al., binding of CheY-P to the FliM-FliN complex induces relative movements of the subunits in the lower part of the C-ring. This causes a movement in the upper part of the C-ring that alters the stator-rotor interaction and changes the rotation direction of the motor [34]. In their view, the switch acts as a mechanical device and converts the small movements induced by CheY-P at the bottom of the C-ring into a large coherent conformational change in the upper part of the ring. Their cross-linking experiment furthermore suggests that the middle domain of the FliM proteins (FliM$_M$) in a CW motor tilts relative to those in a CCW motor, causing a shift at the FliM$_M$-FliM$_M$ interface. Cooperativity could arise within the switch from such conformational change. Studies also showed that the FliM, FliN, and FliG proteins that form the C-ring are conserved in a wide variety of species, including *V. alginolyticus* [35, 36]. Although CheY-P affects the switching behavior of *V. alginolyticus* differently from *E. coli*, it is reasonable to assume that the switching mechanism described above is general and applicable to *V. alginolyticus*.

Below we propose a *minimal* model aimed at mimicking $P(\Delta_f)$ and $P(\Delta_b)$ seen in our experiment. Because the role of CheY-P on the motor is more-or-less symmetric for the two motor states [32], in the ensuing discussion it suffices to consider only one of these transitions, say from CCW to CW. We assign the conformation of the subunit to be active (inactive) when the motor is in CW (CCW) direction. Due to FliM$_M$-FliM$_M$ interaction, the active subunits could form a continuous domain of size $n$, and the motor switches whenever $n$ exceeds a critical number $N_C$.

A simple probabilistic description of finding $n$ active subunits on the switch ring at time $t$ is the master equation, $\frac{dp_n(t)}{dt} = k_-(n+1)p_{n+1}(t) + k_+(n-1)p_{n-1}(t) - (k_+(n) + k_-(n))p_n(t)$, where $p_n(t)$ is such probability, $k_+(n)$ and $k_-(n)$ are rates of $n$ being increased or decreased by unity. Since binding of CheY-P facilitates exiting the current motor state as deduced from our observation, we assume that $k_+ > k_-$ and they are constants. This could happen when the growth of an active domain is enhanced by binding of CheY-P. In the continuum (or large $N_C$) limit, the master equation describes a biophysical process of "diffusion" with a drift

$$\frac{\partial p(x,t)}{\partial t} + V\frac{\partial p(x,t)}{\partial x} = D\frac{\partial^2 p(x,t)}{\partial x^2}, \tag{3}$$

where $x = n/N_C$, $D = (k_+ + k_-)/2N_C^2$, and $V = (k_+ - k_-)/N_C$. The fluctuation in the domain size $n$ is thus equivalent to the motion of a driven Brownian particle in a one-dimensional space, and its first-passage time distribution can be calculated by assuming that a particle is released at $x = 0$ when $t = 0$, and one measures its transmission probability at $x = 1$ as a function of $t$ [37, 38, 39]. The equivalence of the two systems allows us to find the dwell-time distributions, $P(\Delta_f)$ and $P(\Delta_b)$, with the result

$$P(\Delta_s) = \left(\frac{t_{Ds}}{2\pi \Delta_s^3}\right)^{1/2} \exp\left[-\frac{(1 - \Delta_s/t_{Ps})^2}{2(\Delta_s/t_{Ds})}\right], \tag{4}$$

where $t_{Ps} = 1/V_s$ and $t_{Ds} = 1/(2D_s)$ with $s \in \{f, b\}$. We note that although the above derivation is for a particular scenario of the flagellar motor switch, it has a general utility for other biophysical processes that are driven by a constant "force" in a noisy environment. The thermodynamic irreversibility arises via the absorbing boundary condition imposed for solving the first-passage problem [39]. Before proceeding further, it is useful to briefly describe mathematical features of Eq. 4: First, it cuts off sharply for small $\Delta_s$ and has an exponential tail for large intervals, both are qualitatively consistent with our observed dwell-time distributions. Second, Eq. 4 is peaked at $\Delta_{smax} = (t_{Ps}/2)\left(\sqrt{(3\gamma)^2 + 4} - 3\gamma\right)$ with the mean and the standard deviation given respectively by $\langle\Delta_s\rangle = t_{Ps}$ and $\sigma_\Delta = \gamma^{1/2}t_{Ps}$, where $\gamma \equiv t_{Ps}/t_{Ds}$.

Despite its simplicity, this model describes our observations *remarkably* well as delineated by the red lines in Figs. 3 and 5. Here for each PDF, $t_P$ and $t_D$ are the only fitting parameters, and their numerical values are listed in Table 1. We noticed that both the diffusion time $t_D$ and the propagation time $t_P$ depend on the state of the motor $s \in \{f, b\}$ as well as the media used. Specifically, it is found that while $t_{Pf} \simeq t_{Pb}$, there is a considerable difference between $t_{Df}$ and $t_{Db}$. Also, the presence of phenol in the medium significantly reduces these constants. The goodness of the fits in Fig. 3 shows that $P(\Delta_f)$ and $P(\Delta_b)$ belong to the same family of functions.

|  | $t_{Ds}$ (s) | $t_{Ps}$ (s) |
| --- | --- | --- |
| | Forward ($s = f$) | |
| TMN medium | $1.06 \pm 0.06$ | $0.50 \pm 0.02$ |
| TMN+10 mM phenol | $0.52 \pm 0.02$ | $0.31 \pm 0.01$ |
| | Backward ($s = b$) | |
| TMN medium | $2.2 \pm 0.1$ | $0.55 \pm 0.01$ |
| TMN+10 mM phenol | $1.79 \pm 0.05$ | $0.34 \pm 0.01$ |

Table 1: Relevant time scales of motor dynamics. The uncertainties of $t_P$ and $t_D$ are calculated from the estimated covariance matrix.

## Summary

In this study we have witnessed a bacterial flagellar motor switch that operates very differently from that of *E. coli*. For *E. coli*, the regulator CheY-P behaves as a CW rotation enhancer; binding of CheY-P increases the transition rate from CCW to CW state but reduces the transition from CW to CCW state [2, 3]. For *V. alginolyticus*, on the other hand, CheY-P behaves as a switching facilitator; binding of CheY-P increases the exiting rate regardless of its current state. We posit that this type of regulation is well suited for bacteria that are capable of bidirectional swimming and chemotaxis [32].

A salient feature of *V. alginolyticus'* polar flagellar motor switch is the presence of a refractory period during which the motor reversal is strongly inhibited. This is also very different from *E. coli* for which upon switching to a new state, it can immediately switch back. Protection of a nascent state is commonly seen in digital electronics. Since high fidelity in execution of a program is so important, the "dead" time after a switch is built into logical gates of a circuit. For marine bacteria that execute the 3-step motility pattern, the "dead" time can be biologically significant. We believe that this is microorganisms' means of combating noise, ensuring that its switching decision is not overwritten by stochastic noise in a short time. This is particularly significant in oceans where nutrients are subject to dispersion by turbulence. We note that despite stochasticity of turbulent fluid flows, dispersion of a scalar quantity in small scales are more-or-less deterministic and obeys the physical law of mixing. The existence of such mixing time allows the bacteria to develop an anticipatory response, which might explain the short-time inhibition of motor switching observed in the marine bacteria.

To illustrate the idea, we take the typical energy dissipation rate of turbulence near the surface layer of ocean to be $\epsilon \simeq 0.1 \, \text{cm}^2/\text{s}^3$ and the viscosity $\nu \simeq 0.01 \, \text{cm}^2/\text{s}$ [40]. An important spatial scale of turbulence is the Kolmogorov scale, $\ell_\eta = (\nu^3/\epsilon)^{1/4}$, which marks the termination of the inertia dominated flow and the beginning of a viscous subrange. For the given $\epsilon$ and $\nu$, we find $\ell_\eta \simeq 0.06 \, \text{cm}$. Marine bacteria live in a world in which the typical length scale they sense is less than $\ell_\eta$. Consider now a nutrient patch that is dispersed by turbulence. If for the scales $\ell < \ell_\eta$ the nutrient is uniformly distributed, the bacteria may just give up chemotaxis because searching has no benefit. However, owing to the molecular diffusivity $D_0$ of small nutrient molecules being typically several thousand times smaller than the kinematic viscosity $\nu$ of sea water, the nutrients are not distributed uniformly, but rather in patches and striations similar to the stirred milk in a coffee mug. Turbulence causes these spatial inhomogeneities to thin and eventually dissolve at a scale $\ell_C = (\nu D_0^2/\epsilon)^{1/4}$, which is known as the Batchelor scale [41]. A back-of-the-envelope calculation for small amino acids, such as serine ($D_0 \simeq 900\,\mu\text{m}^2/\text{s}$), shows $\ell_C \simeq 17\,\mu\text{m}$. Thus over a range of spatial scales $\ell_C < \ell < \ell_\eta$ (or $20 < \ell < 600\,\mu\text{m}$ for the present case), known as the viscous-diffusion subrange, the marine bacteria can benefit from non-uniform distribution of nutrients if an appropriate chemotactic strategy is employed. We note that since $\ell_C \propto \epsilon^{-1/4}$, the higher the turbulence intensity the smaller the dissolving scale $\ell_C$. Moreover, because of the small (1/4) exponent, the $\epsilon$ dependence is weak, and we expect that $\ell_C \simeq 20\,\mu\text{m}$ should not change much under different conditions. Thus, it is reasonable that for a bacterium to follow changes in a nutrient field, it has to swim the minimal distance $\ell_C$ because otherwise the chemical landscape is featureless. Because the typical swimming speed of a marine bacterium is $v_{sw} \simeq 100\,\mu\text{m/s}$ [42], it follows that the persistent swimming time should be $\sim 0.2$ s. This agrees rather well with the peak positions of $P(\Delta_f)$ and $P(\Delta_b)$ seen in our experiment. The biological and ecological implication of the above observation is significant and should be studied in future experiments.

# Materials and Methods

## (A) Dwell Time and Uncertainty Determination

The marine bacterium *V. alginolyticus* is a 3-step swimmer with a unique motility pattern; the recorded bacterial trajectories consist of a distinctive pattern of run, reverse, and flick, allowing identification of bacterial orientation. Even though the run (f) and reversal (b) intervals are stochastic, the cyclic 3-step pattern is distinct, facilitating tracking and identification of individual motor reversal events.

**Tracking an Ensemble of Bacteria**

Videos of free swimming bacteria YM4 in a 10-$\mu$m deep chamber were taken at the video speed of 30 fps using a 60× objective and a CCD camera (Hamamatzu, EM-CCD C9100). The image size is 512×512 pixels and each pixel measures $0.25 \times 0.25\,\mu\text{m}^2$, which is close to the diffraction limit of optical microscopy. For a non-swimming *V. alginolyticus*, due to thermal diffusion, the displacement along its cell-body axis is $\sim 0.14\,\mu$m and the angular deviation from the cell body's semi-major axis is $\sim 0.11$ rad in 33 ms (see SI). We therefore set up a conservative criterion that during a motor reversal, if the displacement of a cell is less than 0.5 $\mu$m and the cell body's orientation changes less than 0.33 rad between the $i$th and the $(i-1)$th frame, the motor state during the $i$th frame is undetermined, and the uncertainty associated with the moment of this reversal increases by ±16.7 ms. Figs. 6(A-D) illustrate typical examples where positions of a cell in 6 consecutive frames were displayed, showing different scenarios of CCW→CW and CW→CCW transitions. Denote the cell's position in the $i$th frame as $\vec{x}_i$ and the displacement $\Delta\vec{x}_i = \vec{x}_i - \vec{x}_{i-1}$. In sequence (A), since $\Delta\vec{x}_4$ and $\Delta\vec{x}_5$ are pointing in opposite directions, the cell changes its swimming direction between the 4th and the 5th frame, giving the moment of motor reversal at 133.3±16.7 ms. Slower responses are occasionally observed as illustrated in Fig. 6(B). Here $|\Delta\vec{x}_4|$ is less than 2 pixels, while $\Delta\vec{x}_3$ and $\Delta\vec{x}_5$ are pointing in opposite directions. In this case the motor reversal moment

is assigned as $116.7 \pm 33$ ms. In the above two cases, the transition is from CCW to CW because the displacement vectors before and after the motor reversal are anti-parallel. We noticed that the duration of the CW→CCW transition is comparable to that of CCW→CW transitions. Sequence (C) illustrates such a transition, where the cell orientation changes by $\sim 3\pi/4$ during the 3rd and 4th frames; the moment of the flick is thus $100 \pm 16.7$ ms. Sequence (D) illustrates a different case where the cell body does not translate noticeably for 2 consecutive frames (66.7-133.3 ms), but its rotation is clearly discernible in the 3rd and 4th frames. The observed angle of rotation is $\sim \pi/6$ and is significantly greater than what would result from thermal diffusion. The transition moment in this case is assigned to be $100 \pm 16.7$ ms. If two adjacent transition moments are determined to be $t_1 \pm \delta_1$ and $t_2 \pm \delta_2$, the dwell time can be calculated $\Delta_s = t_2 - t_1$ with the uncertainty $\sigma_s = \sqrt{\delta_1^2 + \delta_2^2}$, where $s \in \{f, b\}$. Using the above method, the smallest swimming interval that can be determined is 33 ms with an uncertainty of $\sqrt{16.7^2 + 16.7^2} = 23.6$ ms. We found that $\sim 3\%$ of $\sigma_f$ and $\sim 4\%$ of $\sigma_b$ are larger than 66 ms (see SI), which sets the limit of the resolution of swimming intervals that can be measured using the video microscopy. We note that the sequences of events recorded in Fig. 6 are in good agreement with those reported in Ref. [19], where a high-speed (1000 fps) imaging method was used.

**Tracking Individual Bacteria**

For long-term observations of a single bacterium, the 10-$\mu$m deep chamber was placed on a motorized stage (SD Instrument) controlled by a joystick to keep the selected cell in the field of view. The trajectory was recorded at 24 fps using a Nikon camera (Nikon D90) under a 20× objective. Due to the low magnification of the objective, a single YM4 cell can be tracked for 10 minutes (more details in SI). As a trade-off, however, the resolution of motor reversal moment is reduced to $\sim 0.1$ s.

## (B) Viscoelastic Response of the Propulsive Apparatus of *V. alginolyticus*

The dwell times $\Delta_f$ and $\Delta_b$ we measured are influenced by the overall response of a cell body to rotational fluctuations of the flagellar motor. It is therefore important to understand how fluctuations at the motor level affect the motion of the cell body and our measurements. Specifically, it would be interesting to know the response times of the propulsive system.

A noteworthy feature of *V. alginolyticus* is that the cell body is propelled by a single flagellum connected to a motor by a straight hook [18]. Without the need to form a bundle, the hook bending stiffness of *V. alginolyticus* is much larger than that of *E. coli*; e.g., $EI$ for *V. alginolyticus* is $\sim 3.6 \times 10^{-26}$ N·m$^2$ whereas $\sim 1.6 \times 10^{-28}$ N·m$^2$ for *E. coli* [19, 43]. The large $EI$ allows fast transmission of mechanical disturbances as the viscoelastic response time is inversely proportional to $EI$. The same can also be said about the filament as it is well known that *E. coli*'s flagellum experiences multiple morphological transformations upon motor reversals and under shear flows [44, 45]. But in *V. alginolyticus*, despite its much larger swimming speed, no morphological transformation was observed and flagellar deformation is very minute upon motor reversals [46]. Polar flagellation moreover allows the force and torque to be transmitted directly to the cell body along the body axis. This feature greatly simplifies the calculation of the response time.

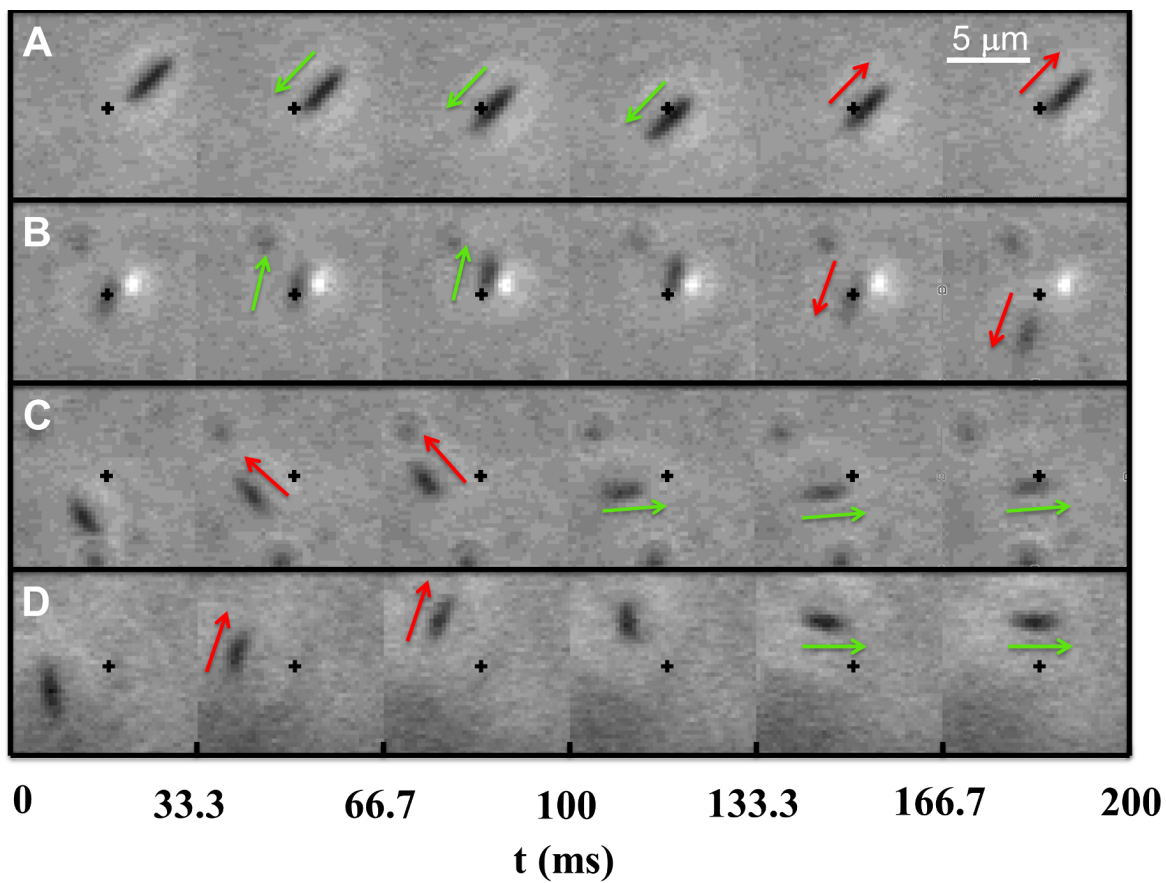

Figure 6: Motor reversal moments determined by video microscopy. Typical motor reversal events are displayed in (A-D), where the green and red arrows indicate the forward and backward swimming directions, and the cross is a stationary reference point. In (A, B), the transition is CCW→CW and in (C, D), the transition is CW→CCW. Note that for CW→CCW transitions, despite a large angular displacement, the translational motion of the cell body can be delayed as delineated by the 3rd and 4th frames in (D). The changes in the swimming directions along with the changes in the cell-body orientation allow the moment of the motor reversal event to be determined.

Below we provide the main result of our analysis while more details can be found in SI. Suppose the flagellar motor switches from CW to CCW rotation at $t = 0$, and we would like to know how long it takes for the cell body to respond to such a switch. The following sequence of events are expected: first the strains that exist on the hook and filament during the CW interval will relax, and then new strains will build up in these components when the motor rotates in the opposite direction. This process can be visualized as elastic waves propagating first along the hook, then the filament, and finally causing the cell body to move. The time scales for these events can be characterized respectively by $\tau_h$, $\tau_f$, and $\tau_b$, yielding the total response time $\tau = \tau_h + \tau_f + \tau_b$. The response time for the cell body, taking into account its inertia, is very small with $\tau_b \sim 10^{-7}$ s. The response times for the hook and the filament are also small, $\sim 10^{-4}$ s or less, when calculated using a linear elastic theory (see SI). Thus, the elastic response time is expected to be $\tau \sim 10^{-4}$ s.

The calculation above is consistent with the experimental finding of Son et al. [19] who used high-speed video imaging microscopy to investigate the mechanisms that causes *V. alginolyticus* to flick upon the motor reversal from CW to CCW (or from backward to forward swimming) reported in Ref. [20]. Recording at 1000 fps, these investigators discovered that when the flagellar motor switches from CW to CCW rotation, initially the cell body back-tracks the backward swimming path for $\sim 10$ ms, it then sharply changes it orientation within another $\sim 10$ ms. The short latent period, right after the motor reversal and just before flicking, is a result of unwinding and then winding of the flagellar hook; i.e., it starts from being taut, to loose, and becomes taut again. Son et al.'s measurement showed that the flexural rigidity $EI$ of the hook increases by an order of magnitude when the hook is loaded as compared to when it is relaxed. The reduced $EI$ makes the hook more pliable, and under compression, it buckles giving rise to a sharp turn of the cell body after a 10 ms latent period. Thus backtracking and abrupt reorientation of the cell body provides a reliable and convenient means for identifying the moment when the flagellar motor reverses, i.e., with a proper instrumentation a change in the cell body movement should be detectable at a time scale $\sim 10^{-4}$ s due to the viscoelastic response to a CCW→CW or CW→CCW transition. Reorientation of the cell-body, or a flick, due to the elastic instability takes a longer time with $\tau_r \sim 10 - 20$ ms, and conservatively we take this slow $\tau_r$ to be the relevant time for determining a motor reversal event using video imaging microscopy. The width of the shaded area in Fig. 3 is two video frames and is about $3\tau_r$.

# Acknowledgments

The *V. alginolyticus* strain YM4 is a kind gift of Prof. M. Homma. We would like to thank Y. Tu for a helpful discussion concerning the modeling of the motor switch. This work is partially supported by the NSF under the grant no. DMR-1305006, and Ms. Xie is supported by the Predoctoral Mellon Fellowship from the University of Pittsburgh.

# References

[1] Berg HC (2003) The rotary motor of bacterial flagella. *Annual Review of Biochemistry* 72:19–54.

[2] Kuo S, Jr. DK (1989) Multiple kinetic states for the flagellar motor switch. *J Bacteriol* 171:6279–6287.

[3] Scharf BE, Fahrner KA, Turner L, Berg HC (1998) Control of direction of flagellar rotation in bacterial chemotaxis. *Proc Natl Acad Sci USA* 95:201–206.

[4] Berg H, Brown D (1974) Chemotaxis in escherichia coli analyzed by three-dimensional tracking. *Antibiot Chemother* 19:55–78.

[5] Bai F, et al. (2010) Conformational spread as a mechanism for cooperativity in the bacterial flagellar switch. *Science* 327:685–689.

[6] Wang F, Yuan J, Berg HC (2014) Switching dynamics of the bacterial flagellar motor near zero load. *Proceedings of the National Academy of Sciences* 111:15752–15755.

[7] Adler J, Dahl M (1967) A method for measuring the motility of bacteria and for comparing random and non-random motility. *J. Gen. Microbiol.l* 46:161–173.

[8] Springer M, Goy M, Adler J (1979) Protein methylation in behavioural control mechanisms and in signal transduction. *Nature* 280:279–284.

[9] Bray D, Levin MD, Morton-Firth CJ (1998) Receptor clustering as a cellular mechanism to control sensitivity. *Nature* 393:85–88 10.1038/30018.

[10] Cluzel P, Surette M, Leibler S (2000) An ultrasensitive bacterial motor revealed by monitoring signaling proteins in single cells. *Science* 287:1652–1655.

[11] Monod J, Wyman J, Changeux JP (1965) On the nature of allosteric transition: a plausible model. *J. Mol. Biol.* 12:88–118.

[12] Koshland D, Nemethy G, Filmer D (1966) Comparison of experimental binding data and theoretical models in proteins containing subunits. *Biochemistry* 5:365–368.

[13] Duke TAJ, Novere NL, Bray D (2001) Conformational spread in a ring of proteins: A stochastic approach to allostery. *J. Mol. Biol.* 308:541–553.

[14] Mochrie S, Mack A, Gegan L (2010) Allosteric conformational spread: exact results using a simple transfer matrix method. *Phys. Rev. E* 82:031913.

[15] Tu Y (2008) The nonequilibrium mechanism for ultrasensitivity in a biological switch: Sensing by maxwell's demons. *Proc Natl Acad Sci USA* 105:11737–11741.

[16] Kawagishi I, Maekawa Y, Atsumi T, Homma M, Imae Y (1995) Isolation of the polar and lateral flagellum-defective mutants in vibrio alginolyticus and identification of their flagellar driving energy sources. *J. Bacteriol.* 177:5158–5160.

[17] Blake PA, Weaver RE, Hollis DG (1980) Diseases of humans (other than cholera) caused by vibrios. *Annual Review of Microbiology* 34:341–367.

[18] Terashima H, Fukuoka H, Yakushi T, Kojima S, Homma M (2006) The vibrio motor proteins, motx and moty, are associated with the basal body of na+-driven flagella and required for stator formation. *Mol Microbiol* 62:1170–1180.

[19] Son K, Guasto JS, Stocker R (2013) Bacteria can exploit a flagellar buckling instability to change direction. *Nat Phys* 9:494–498.

[20] Xie L, Altindal T, Chattopadhyay S, Wu X (2011) Bacterial flagellum as a propeller and as a rudder for efficient chemotaxis. *Proc Natl Acad Sci USA* 108:2246–2251.

[21] Purcell E (1977) Life at low reynolds number. *Am. J. Phys.* 45:3–11.

[22] Colquhoun D, Hawkes A (1981) On the stochastic properties of single ion channels. *Proc. R. Soc. London* 211:205–235.

[23] Purcell EM (1956) The question of correlation between photons in coherent light rays. *Nature* 178:1449–1450 10.1038/1781449a0.

[24] Glauber R (1965) *in Quantum Optics and Electronics (eds C. DeWett, A. Blandin, and C. Cohen-Tannoudji)* (Gordon and Breach, New York), pp 63–185.

[25] Fienberg SE (1974) A biometrics invited paper. stochastic models for single neuron firing trains: A survey. *Biometrics* 30:399–427.

[26] Berry MJ, Meister M (1998) Refractoriness and neural precision. *The Journal of Neuroscience* 18:2200–2211.

[27] Turner L, Caplan SR, Berg HC (1996) Temperature-induced switching of the bacterial flagellar motor. *Biophys J* 71:2227–2233.

[28] Alexander RP, Lowenthal AC, Harshey RM, Ottemann KM (2010) Chev: Chew-like coupling proteins at the core of the chemotaxis signaling network. *Trends in Microbiology* 18:494–503.

[29] Kojima M, Kubo R, Yakushi T, Homma M, Kawagishi I (2007) The bidirectional polar and unidirectional lateral flagellar motor of vibrio alginolyticus are controlled by a single chey species. *Molecular Microbiology* 64:57–67.

[30] Homma M, Oota H, Kojima S, Kawagishi I, Imae Y (1996) Chemotactic responses to an attractant and a repellet by the polar and lateral flagellar systems of vibrio alginolyticus. *Microbiology* 142:2777–2783.

[31] Yamamoto K, Macnab RM, Imae Y (1990) Repellent response functions of the trg and tap chemoreceptors of escherichia coli. *J Bacteriol* 172:383–388.

[32] Xie L, Lu C, Wu XL (2010) Marine bacterial chemoresponse to a stepwise chemoattractant stimulus. *Biophysical Journal* 108:766–774.

[33] Altindal T, Chattopadhyay S, Wu X (2011) Bacterial chemotaxis in an optical trap. *PLoS ONE* 6:e18231.

[34] Paul K, Brunstetter D, Titen S, Blair DF (2011) A molecular mechanism of direction switching in the flagellar motor of escherichia coli. *Proc Natl Acad Sci USA* 108:17171–17176.

[35] Chen S, et al. (2011) Structural diversity of bacterial flagellar motors. 30:2972–2981.

[36] Li N, Kojima S, Homma M (2011) Sodium-driven motor of the polar flagellum in marine bacteria vibrio. *Genes to Cells* 16:985–999.

[37] Darling D (1953) *The first passage problem for a continous Markoff process* (Rand Corporation, Pennsylvania).

[38] Naber H (1996) Two alternative models for spontaneous flagellar motor switching in halobacterium salinarium. *J. Theor. Biol.* 181:343–358.

[39] Schrodinger E (1915) Zur theorie der fall- und steigversuche an teilchen mit brownscher bewegung. *Physikalische Zeitschrift* 16:289–295.

[40] Luchsinger R, Bergersen B, Mitchell J (1999) Bacterial swimming strategies and turbulence. *Biophys. J.* 77:2377–2386.

[41] Batchelor B (1959) Small-scale variation of convected quantities like temperature in turbulent fluid, par1. general discussion and the case of small conductivity. *J. Fluid Mech.* 5:113–133.

[42] Mitchell J (1991) The influence of cell size on marine bacterial motility and energetics. *Microb. Ecol.* 22:227–238.

[43] Sen A, Nandy RK, Ghosh AN (2004) Elasticity of flagellar hooks. *Journal of Electron Microscopy* 53:305–309.

[44] Turner L, Ryu WS, Berg HC (2000) Real-time imaging of fluorescent flagellar filaments. *J Bacteriol* 182:2793–2801.

[45] Hotani H (1982) Micro-video study of moving bacterial flagellar filaments: Iii. cyclic transformation induced by mechanical force. *Journal of Molecular Biology* 156:791–806.

[46] Takano Y, Yoshida K, Kudo S, Nishitoba M, Magariyama Y (2003) Analysis of small deformation of helical flagellum of swimming vibrio alginolyticus. *JSME Int. J. Ser. C.* 46:1241–1247.

# An Element of Determinism in a Stochastic Flagellar Motor Switch
# Supporting Information

## Effect of Correlation Between Adjacent Intervals

When there is no correlation between switching events that are governed by the equilibrium model, the autocorrelation function $C_I(t)$ decays monotonically. However, small but discernible correlations between adjacent intervals have been observed in cells #1 and #4 with $C_\Delta(1) = 0.11$, while the standard deviation of $C_\Delta(m)$ is 0.03-0.04 for $m > 1$ (see Eq. 2 for definition of $C_\Delta(m)$). Although such correlation cannot account for $F < 1$, it can cause $C_I(t)$ to oscillate. To test this possibility, we shuffled the intervals in different ways to evaluate the effect of temporal correlation of switching events on $C_I(t)$. The time series for the #4 cell is chosen because the corresponding $C_I(t)$ has the strongest oscillation. First, to eliminate the correlation between adjacent intervals, forward and the backward intervals $\Delta_f$ and $\Delta_b$ are shuffled among themselves to generate twenty randomized binary sequences $I'(t')$. Autocorrelation functions were then computed and averaged, yielding $C_I'(t)$ that is displayed by the indigo curve in Fig. S1. As can be seen, even in the absence of adjacent-interval correlation, $C_I'(t)$ still oscillates. However, $C_I'(t)$ does deviate from $C_I(t)$ noticeably and the deviation is consistent with the result $C_\Delta(1) = 0.11$. We next shuffled time-ordered pairs $(\Delta_f,\Delta_b)$ with each other to obtain another binary sequence $I''(t')$. This procedure maintains the correlation between $\Delta_f$ and $\Delta_b$ within a swimming cycle but the long-time correlation is destroyed. The average autocorrelation functions $C_I''(t)$ resulting from twenty such shuffles were shown by the brown curve. As seen $C_I''(t)$ is nearly identical to $C_I(t)$, suggesting that there is very little correlation between pairs of $(\Delta_f,\Delta_b)$.

Another way to demonstrate the lack of long-time correlation is to calculate the distribution of $C_\Delta(m)$ after shuffling the forward and the backward intervals $\Delta_f$ and $\Delta_b$ among themselves, where $1 < m \leq 200$. It is evident in Fig. S2 that shuffling has little effect on the PDFs of $C_\Delta(m)$.

The above statistical analyses allow us to conclude that even though temporal correlation is discernible, these correlations exist only between adjacent intervals. The observed $F < 1$ and oscillations in $C_I(t)$ must be due to the non-monotonic distributions of $\Delta_f$ and $\Delta_b$ that show prominent peaks at finite times.

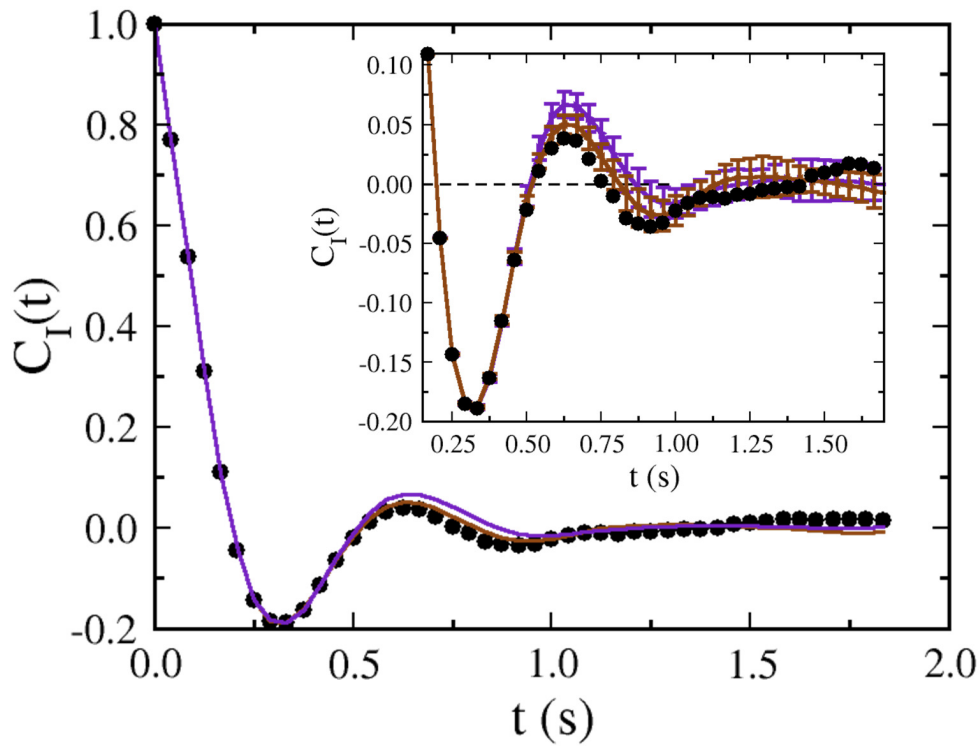

Figure S1: Autocorrelation functions of bacterial swimming intervals. $C_I(t)$ computed from the time series of the cell #4 is depicted by the black dots. The autocorrelation functions $C'_I(t)$ and $C''_I(t)$ are shown by the indigo and brown curves (see main text). To aid visualization, the region $0.2 < t < 1.5$ s was amplified in the inset. Here, the error bars for $C'_I(t)$ and $C''_I(t)$ represent the standard deviation resulting from twenty realizations of randomly shuffled time sequences.

## SI Materials and Methods

### Bacterial Strains and Cultures

The *V. alginolyticus* strain YM4 (Pof$^+$Laf$^-$) is a gift of Professor M. Homma [1]. The bacteria for video microscopy were grown in a minimal medium [2] (0.3 M NaCl, 10 mM KCl, 2 mM $K_2HPO_4$, 0.01 mM $FeSO_4$, 15 mM $(NH_4)_2SO_4$, 5 mM $MgSO_4$, 1% glycerol, and 50 mM Tris-HCl (pH 7.5)) to an optical density 0.2-0.3 at 30 °C. 1.5 mL culture was harvested and spun down at 2000×g for 3 minutes. After removing the supernatant, 1 mL TMN motility medium (50 mM Tris-HCl (pH 7.5), 5 mM $MgCl_2$, 5 mM glucose, 30 mM NaCl, and 270 mM KCl) was used to resuspend the culture followed by a 5-minute centrifuging at 500×g [1]. 300-400 μL supernatant was then carefully diluted into 2 mL TMN and shaken at 200 rpm at room temperature for at least half an hour before observation.

The bacteria for optical trapping were grown overnight in 2 mL VC medium (0.5% polypeptone, 0.5% yeast extract, 0.4% $K_2HPO_4$, 3% NaCl and 0.2% glucose) at 30 °C while shaken at 200 rpm. The overnight culture was diluted 1:100 into VPG (1% polypeptone, 0.4% $K_2HPO_4$, 3% NaCl and 0.5% glycerol) and grown for 3-4 hours at 30 °C while shaken at 200 rpm [3]. The cells were then washed twice with TMN medium at 900×g for 2 minutes before resuspended in TMN and incubated for 8 hours before measurement.

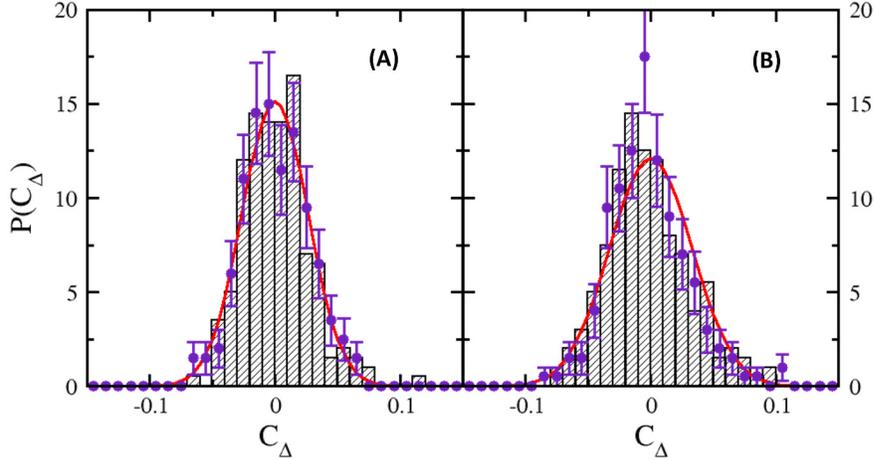

Figure S2: PDF of $C_\Delta(m)$ with and without shuffling. (A) The PDFs of $C_\Delta(m)$ calculated using the original time sequence (bars) and the shuffled sequence (indigo dots) obtained from cell #4, where $1 < m \leq 200$. The red curve is the same normal distribution shown on the right panel of Fig. 2(A). (B) The same quantities obtained using cell #5.

## Video Tracking of Individual Swimming Bacteria

We took videos of *V. alginolyticus* swimming using a 20× objective (Nikon, Plane Fluor 20× N.A.=0.45) in the phase contrast mode and a Nikon D90 camera at 24 fps. The observation chamber is purchased from Hawksley (Z3BC1B) and has a depth of 10 $\mu$m. The shallow depth of the chamber and the use of the low magnification allow a relatively long term recording of individual cells in the field of view. In order to follow individual *V. alginolyticus* cells for a long time, $\sim$ 10 minutes, the observation chamber was placed on a motorized stage (SD instrument, MC2000 controller, 200 Cri motorized linear stage) controlled by a joystick. The stage was moved to keep the cell inside the field of view. These videos were analyzed using the ImageJ manual tracking plug-in. One typical bacterial trajectory is displayed in Fig. S3, showing distinctively different swimming segments during forward (green) and backward (red) intervals. Due to hydrodynamic interactions with boundaries, trajectories are usually curved [4, 5]. As observed in Fig. S3, the forward swimming segment is curved in the CCW direction but the backward swimming segment is curved in the CW direction. Observations also show that the backward segments curve more strongly than the forward ones [4]. Moreover, when a cell switches from forward to backward swimming, the cell body's orientation is more or less the same. On the other hand, when a cell switches from backward to forward swimming, it usually flicks. During the flicking, the cell slows down and there are abrupt changes either in the cell body orientation, the shape, or both. Using these criteria, most motor reversal events can be determined as CCW→CW or CW→CCW without ambiguity. Those reversal events that are difficult to determine can then be identified based on the fact that the motor alternates between CCW and CW rotations. As the trade off of the long observation time, the temporal resolution of the switching moments is not as good as when the videos are taken at 30 frames per second using a 60× objective. These long bacterial trajectories are suitable for analysis such as estimating the Fano factor and temporal correlation where uncertainties are averaged out, and they were not used for calculating $P(\Delta_f)$ and $P(\Delta_b)$.

An often used method in *E. coli* studies is the rotation assay, which relies on tethering the cell body or a flagellum to a surface. Although convenient, the method can skew motor switching behavior as recent studies indicated that switching statistics are influenced by the load [6, 7]. The use of freely swimming bacteria in this study, although tedious, is free of these complications, and the measured switching statistics directly reflect the unperturbed physiological state of the bacteria.

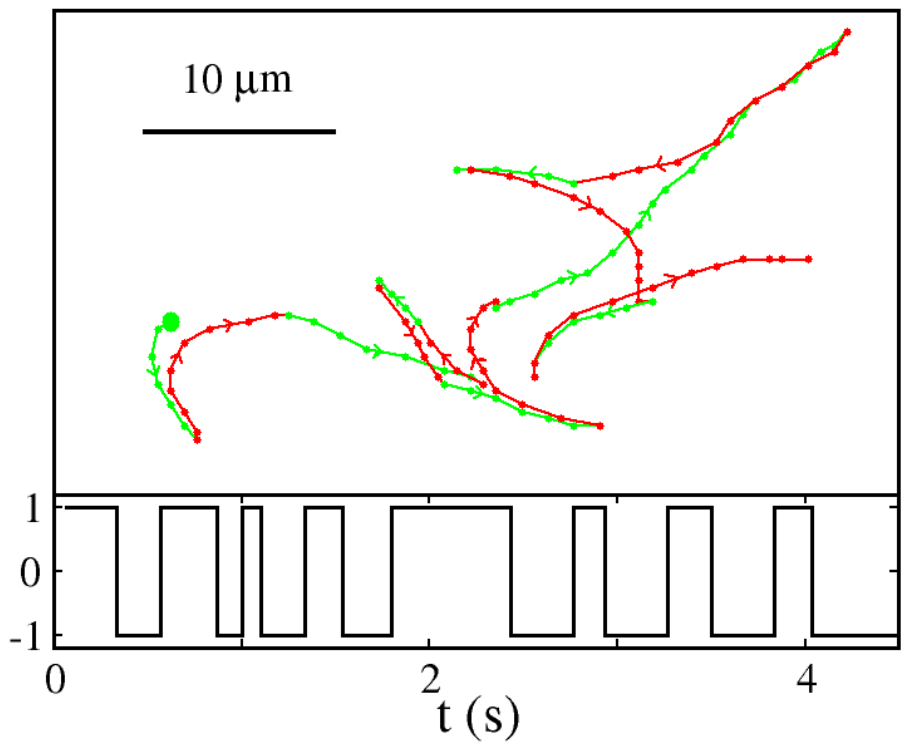

Figure S3: A typical bacterial trajectory and its binary presentation $I(t')$. The green and red lines denote the forward and the backward swimming segments, respectively. The large green dot indicates the starting point of the trajectory at $t = 0$ and the small dots are the positions of the bacterium at an equal time interval $\Delta t = 0.042\,\text{s}$. The arrows indicate the swimming direction. To aid visualization, some dots are shifted slightly to avoid overlapping. On the bottom of the figure the bacterial trajectory is binarized according to $I(t') = +1$ if the cell swims forward and $I(t') = -1$ if it swims backward.

# Recording Motor Switching Events of Individual Bacteria Using Optical Trapping

A home-built optical trap was used to detect motor reversals of individual cells as described in Ref. [8]; here only the relevant aspects are delineated. Using radiation pressure from a tightly focused laser beam ($\lambda = 1054$ nm, $\sim 50$ mW), the elliptically shaped bacterium is trapped along the optical axis as shown in Fig. 4(A, B). The trapped bacterium has a limited range of movement ($\sim 1$ $\mu$m) along the optical axis, but its rotational degree of freedom about this axis is unrestricted. Because the flagellum and the cell-body axis is rarely perfectly aligned, a swimming bacterium wobbles in the optical trap and this small irregular motion can be recorded using a two-dimensional position sensitive detector (PSD), resulting in a time series $(x(t), y(t))$ [9]. In our previous work, we demonstrated that by a simple Fourier transformation of $(x(t), y(t))$, the rotation frequencies of the cell body, the flagellum, and the moments of motor reversals can be determined [9, 8].

For the present work, a similar but simpler approach is taken to identify the motor reversals. Specifically we found that by tilting the optical ($z$) axis of the trap, the bacterial motion along the $z$ axis has a projection on the $x$ axis and can be recorded by the PSD at a sampling rate of 10 kHz. In the experiment, each cell is trapped for several seconds to obtain the time trace $x(t)$. To avoid potential artifacts, as a result of photodamage to the cells, only the first 3 s of data is processed. A typical time trace is displayed in Fig. S4 (also Fig. 4(C) in the main text). The histogram constructed from $x(t)$ can be fitted to a sum of two Gaussian functions as delineated in the right panel of Fig. S4. The centers of these two Guassians, $+x_0$ and $-x_0$, correspond to the two stable positions of the bacterial cell body in the optical trap. Since the standard deviation of the Gaussian functions are about $x_0/3$, $+x_0$ and $-x_0$ are well separated despite the noise in the measurements. This allows individual motor reversals to be determined by a computer with little ambiguity.

To begin with, we measured $P(t_s)$, the PDF of $\sim 180$ transition times between $+x_0$ and $-x_0$. As seen in Fig. 4(D), $P(t_s)$ is peaked at 15 ms and has a broad tail, yielding the mean switching time $\bar{t}_s \simeq 22$ ms. As discussed in the main text, this time corresponds to a bacterium moving $\sim 1$ $\mu$m at the swimming speed of $v_{sm} \simeq 55$ $\mu$m in the trap and can be considered as the temporal resolution of the technique. Next we determined dwell times $\Delta_u$ from a time trace using a Matlab code. First, the transitions between $+x_0$ and $-x_0$ are accentuated by convoluting $x(t)$ with a smooth-derivative filter $F(t) = -\frac{t}{2c^2} \exp{(-t^2/2c^2)}$. Here $c$ sets the time scale by which the raw data is smoothed and the time derivative is taken. Also, $F(t)$ is properly normalized so that when convoluted with the Heaviside function $H(t)$ it yields $\int_{-\infty}^{\infty} H(t - t') F(t') dt' = 1/2$ for $t = 0$.

In Fig. S4, the convoluted data $\Delta x(t) = \int_{-\infty}^{\infty} x(t - t') F(t') dt'$ is displayed by the red curve, where $c = 30$ ms. The prominent maxima and minima of the curve clearly mark the moments when the motor reverses. Second, since there are also secondary maxima and minima in $\Delta x(t)$, their selection or discrimination must be decided. This is carried out by setting the thresholds $x_{th} = \pm k x_0$ with $0 < k < 1$. If a maximum or minimum of $\Delta x(t)$ exceeds the threshold, the corresponding fluctuation in $x(t)$ is counted as a switching event. Otherwise it is discriminated as noise or an incomplete switch. The blue lines in Fig. S4 mark $x_{th}$ with $k = 75\%$, and they clearly discriminate the prominent maxima and minima in $\Delta x(t)$ from the secondary ones. In general, the larger the $k$ (or $x_{th}$) value, the more the secondary maxima and minima are discriminated.

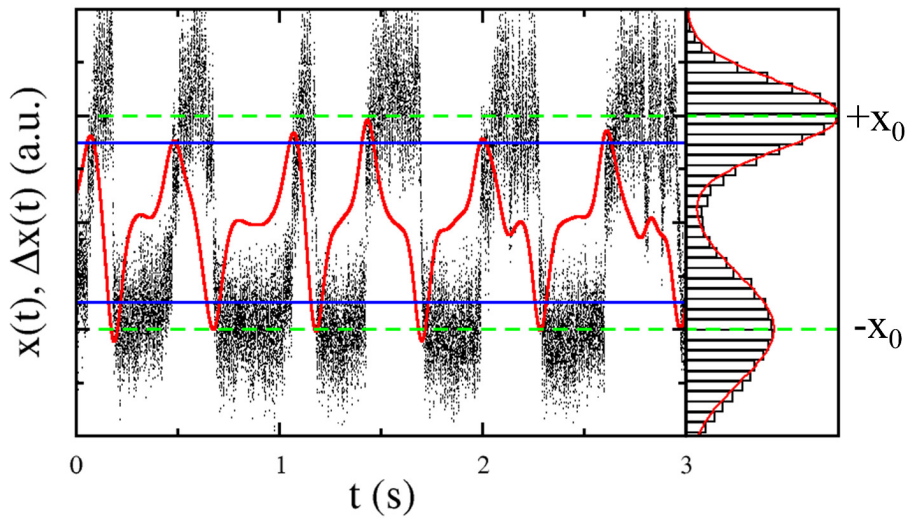

Figure S4: Determination of motor reversal moments from the time trace $x(t)$. The black dots, $x(t)$, are the $x$-projection of the position of a bacterium in the optical trap at every 0.1 ms. The forward and backward swimmings of the bacterium result in two stable positions of the cell body in the trap. Since the optical axis of the trap is slightly tilted, $x(t)$ fluctuates around two constant values, $+x_0$ and $-x_0$, which are marked by the green dashed lines. $x_0$ can be determined precisely by fitting the histogram of $x(t)$ to a sum of two Gaussian functions as delineated by the red curve in the right panel next to the time trace. To determine the moments when the motor changes its rotation state, $x(t)$ is convoluted with $F(t)$, resulting in $\Delta x(t)$, which is displayed by the red curve superimposed over the time trace. As can be seen, the major maxima and mimima match very well with the transitions of $x(t)$, and they exceed the threshold values, $x_{th} = \pm k x_0$, marked by the blue lines.

We also investigated systematically how varying $c$ and $k$ affects the measured dwell-time distribution (or histogram) in the optical trap, and the result is shown Fig. S5. The panels on the top and bottom rows are obtained using $k = 75\%$ and $k = 50\%$, respectively. The three columns, from left to right, correspond to $c = 30$, 20, and 10 ms, respectively. As can be seen, reducing $c$ or $k$ has the same effect of introducing more short intervals into the histogram. Importantly, however, all the histograms remain non-monotonic with a prominent peak at $\Delta_u \simeq 0.2 - 0.3$ s. Taking into consideration the temporal resolution of this technique, which is $\bar{t}_s \simeq 22$ ms, $c$ should be comparable, or slightly greater than $\bar{t}_s$. As seen in Figs. S5(A, D), when $c = 30$ ms, both histograms obtained using $k = 75\%$ and $k = 50\%$ are not significantly different from each other, or for that matter different from the one measured using the video imaging technique.

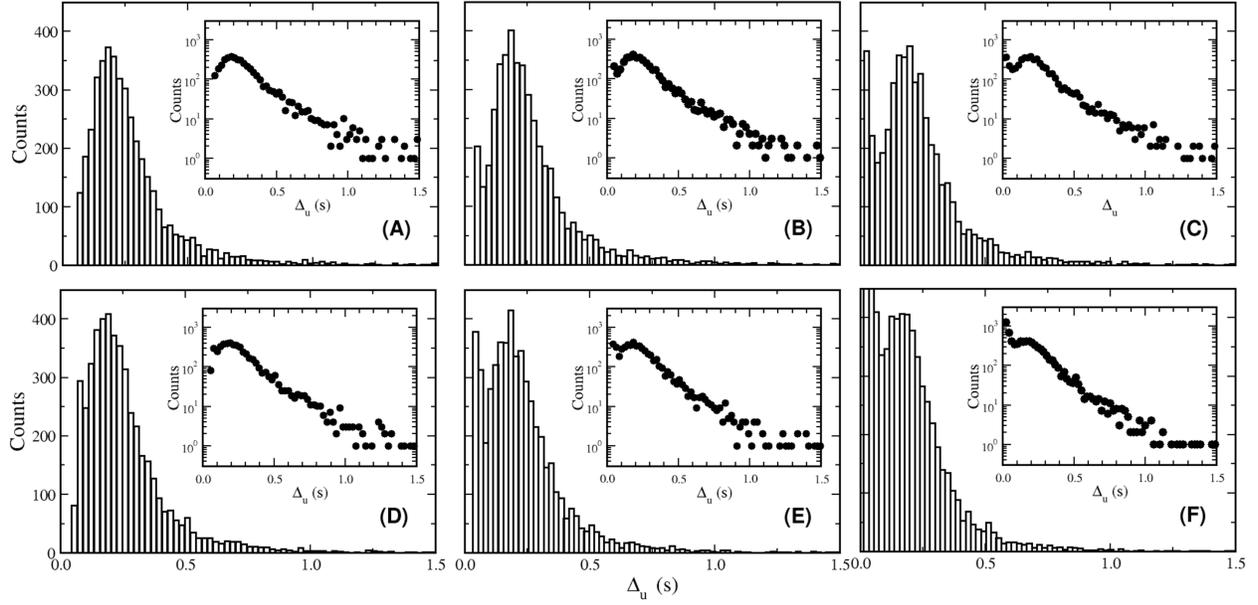

Figure S5: Effects of thresholding and filtering on dwell-time $\Delta_u$ histogram. For each histogram, $\Delta_u$ were determined as described in the text using different values of $c$ and $k$. The combinations of three low-pass filters, $c = 30$ (left column), 20 (middle column), and 10 ms (right column), and two thresholds, $k = 75\%$ (top panels (A-C)) and 50% (bottom panels (D-F)), were tested. It is seen that more short intervals, or incomplete switching events, contribute to the histograms as $c$ or $k$ decreases. Nevertheless, even when $c = 10$ ms $< \bar{t}_s$, the histograms still decay non-monotonically.

## Viscoelastic Responses Times of Cell Body to Motor Reversals

Most motile bacteria use rotating helices, the flagella, to propel themselves in fluids. Although the physical principle of propulsion appears to be universal, its implementation varies considerably among different bacteria. One observes that some bacteria are peritrichously flagellated, such as *E. coli*, *B. subtilis* and *S. typhimurium*; some are monotrichously flagellated, such as *V. alginolyticus*, *C. crescentus*, and *P. haloplanktis*; some are lophotrichously flagellated, such as *V. fischeri* and *Geobacter*; and still others having their flagella embedded in the cell body, such as spirochaetes (*Treponema*, *Borrelia*, *Leptospira*). In the latter case, the whole cell body undulates as the flagella rotates inside, shedding circular waves from one end of body to the other.

While it is not known why and how these different flagellation patterns arise, it is certain that they are the remarkable results of evolution by natural selection. The selective force appears to work its way to minute details, such as the size and the shape of a hook and a flagellum (filament), and their mechanical properties. Take for example the commonly studied bacteria *E. coli* and *V. alginolyticus*. Because of their different flagellation patterns, there is a fundamental difference in the manner by which force and torque are transmitted to the cell body in these bacteria. For *E. coli*, multiple flagella are connected to motors at their bases via elastic hooks that *bend* by 90 degrees [10]. The hook as well as the flagella are relatively soft compared to that of *V. alginolyticus* [11, 12], allowing multiple flagella to coalesce into a coherent bundle during runs and to disperse during tumbles. Due to multiple motors and flagella involved, bundle formation and dissociation involve complicated dynamics, making it difficult to study motor fluctuations using free-swimming cells. Indeed, casual observations of *E. coli* swimming show that run-to-tumble and tumble-to-run transitions are not very sharp. To study motor fluctuations in *E. coli*, scientists monitor rotations of a small bead that is tethered to a flagellar hook in the absence of the filament [7, 13].

In contrast, the marine bacterium *V. alginolyticus* has a polar flagellum that is connected to the motor at its base by a *straight* hook [10]. In this simple body layout, the flagellum and cell body are aligned so that the force and the torque are transmitted along the cell-body axis. Aided by relatively large flexural rigidities $EI$ of the hook and the filament, our calculation below shows that the response time of the cell body to flagellar motor speed fluctuations is very short, $< 10^{-4}$ s, consistent with our observations and others [14].

The accuracy of measuring dwell-time distributions, $P(\Delta_f)$ and $P(\Delta_b)$, depends on how well the interval times $\Delta_f$ and $\Delta_b$ can be determined. This in turn depends on how precisely one can determine the moment when a flagellar motor changes its direction. Here two limiting issues need to be considered: (i) the limit set by the measurements, and (ii) the limit set by the intrinsic response time of the cell body to a change in the motor speed. As (i) has already been addressed in the main text, here only (ii) will be analyzed. Treating the flagellar hook and the filament as elastic elements, one can calculate how a local strain induced by the motor propagates along the hook-filament complex and finally causes the cell body to react. We denote the response time by the hook, the filament, and the cell body by $\tau_h$, $\tau_f$, and $\tau_b$, respectively, and the total response time is a sum of them.

(a) Estimate $\tau_f$: Based on the dark-field microscopic measurement by Nishitoba et al. [15], Takano and his coworkers analyzed small deformations of *V. alginolyticus*' flagellum using the linear elastic theory of Kirchhoff [16]. By comparing the numerical result with the measurement, they estimated the flexural stiffness of flagellar filament to be $EI \sim 10 - 15\,\text{pN} \cdot \mu\text{m}^2$. During the forward or backward swimming, the viscous force per length along the azimuthal $\theta$ direction is $f_\theta \approx 2\,\text{pN}/\mu\text{m}$ [16]. Nishitoba et al.'s measurement also showed that when the helix flagellum is pushing or pulling the cell body, the helix tightens or loosens slightly giving rise to a change in the number of turns by ~0.08 in both directions [16]. This allows us to estimate the torsional stiffness $GJ$ of the filament. For the flagellar filament, the angular displacement $\theta$ and the force $f_\theta$ are related by the Hooke's law,

$$f_\theta L_f R_f = -GJ \frac{\theta - \theta_0}{L_f},$$

where $R_f$ is the radius of the helix, $L_f$ is the contour length of the flagellum, and $\theta - \theta_0 = 0.08 \times 2\pi$ is the twist angle. Using flagellar geometric parameters of *V. alginolyticus* [16], $R_f = 0.23\,\mu\text{m}$ and $L_f = 5.5\,\mu\text{m}$, we found the torsional stiffness $GJ = 28\,\text{pN}\cdot\mu\text{m}^2$. For an elastic coil embedded in a viscous fluid, a local twist $\theta$ will relax, and the equation of motion of $\theta$ is determined by balancing the elastic torques with the viscous ones. This yields

$$C_t R_f \sqrt{R_f^2 + (\frac{\lambda}{2\pi})^2} \partial_t \theta = GJ \frac{\partial^2 \theta}{\partial s^2},$$

where $C_t$ is the tangential dragging coefficient per unit length, $\lambda$ is the pitch, and $s$ is the distance along the contour of the helix. Thus a local twist at one end of the flagellum transmitted to the other end diffusively with a time scale $\tau_f = L^2 C_t R \sqrt{R^2 + (\frac{\lambda}{2\pi})^2}/GJ$. Using the measured $\lambda = 1.27\,\mu\text{m}$ and the calculated drag coefficient $C_t = \frac{4\pi\eta}{\ln(2q/r_f)+1/2}$, where $\eta = 0.01$ cP is the viscosity of water, $r_f = 16$ nm is the radius of the filament [17, 16], and $q = 0.09\Lambda$ with $\Lambda = 1.57\,\mu\text{m}$ being the pitch along the contour of the flagellum [18], we found $\tau_f \simeq 3 \times 10^{-4}\,\text{s}$. Likewise, one can also estimate the compressional relaxation time of the helical coil, yielding nearly the same result. We note that this estimation is consistent with the numerical simulation of Vogel and Stark [19].

(b) Estimate $\tau_h$: Repeat the above linear elastic theory for the hook, the characteristic time is found to be $\propto 4\pi r_h^2 L_h^2 \eta / GJ$, where the radius and length of the hook are $r_h \simeq 0.01\,\mu\text{m}$ and $L_h \simeq 0.08\,\mu\text{m}$, respectively [10]. Although there is no direct measurement, if we assume that the hook is homogeneous and isotropic, it can be estimated that for the hook, $GJ = EI/(1+\nu) \simeq 2.7 \times 10^{-2}\,\text{pN} \cdot \mu\text{m}^2$, where $EI = 3.6 \times 10^{-2}\,\text{pN}\cdot\mu\text{m}^2$ and the Poisson ratio $\nu = 1/3$ [14]. The elastic relaxation time estimated in this way, $3 \times 10^{-7}\,\text{s}$, is very small and cannot be relevant to our experiment.

On the other hand, experiments using *E. coli* cells reveal that the torsional spring constant of the flagellar hook exhibits strong nonlinearity [11]. Using an optical tweezers to wind and unwind a hook that is connected to a locked flagellar motor, Block et al. discovered that the torsional spring constant is ~0.4 pN $\mu$m /rad up to about $\phi_c \simeq 100^o$ of twist, and it then becomes more than an order of magnitude stiffer. In other words, for *E. coli*, once the motor twists the hook over $\phi_c$ the hook can be considered rigid. Since the length of the flagellar hook of *V. alginolyticus* is $\sim 80$ nm while that of *E. coli* is $\sim 50$ nm [20, 21, 10] for *V. alginolyticus*, $\phi_c$ is expected to be correspondingly greater, or about half turn (if the rigidity of the hooks of *E. coli* and *V. alginolyticus* are the same). *V. alginolyticus* uses a sodium motor to power its flagellum, and in the steady state, the motor rotates at an angular frequency of $\sim 600$ Hz when the motility buffer contains 30 mM NaCl [22, 23]. However when motor reverses its direction, the rotation speed is not constant but increases in an almost linear fashion [14]. Thus, the average speed may be taken as $\bar{f} \sim 300$ Hz. This yields a rough estimate of $\tau_h (\equiv \phi_c/\bar{f}) \simeq 0.5/300\,\text{Hz} \simeq 2$ ms. This is a conservative estimate in the sense that while a twist density is introduced at the base of the hook by the flagellar motor, it is released at the distal end. As a result $\tau_h$ is expected to be somewhat longer. A recent experiment using fast video imaging shows that upon a motor reversal from CW to CCW rotation, or a transition from backward to forward swimming, the cell body backtracks for $\sim 10$ ms before it is deflected to a new direction. Interestingly the backtracking can be resolved frame by frame at an interval of 1 ms, suggesting that even during this unwinding period the displacement of the cell-body follows closely the rotation of the motor [14]. The sudden change in the swimming direction, which we termed a flick [24], was interpreted as a buckling instability when the rigidity of the hook is at its lowest and can be associated with the loading time of the hook which is about $10 - 20$ ms.

(c) Estimate $\tau_b$: Here we approximate the cell as an ellipsoid with a semi-major axis $a \simeq 1.5\,\mu$m and a semi-minor axis $b \simeq 0.5\,\mu$m. The translational diffusion coefficient $D_1$ along the cell-body semi-major axis is given by [25],

$$D_1 = \frac{k_B T}{4\pi \eta a}(\ln \frac{2a}{b} - \frac{1}{2}) \qquad (1)$$

where $k_B$ is the Boltzmann constant and $T$ is the temperature. This gives $D_1 \simeq 2.8 \times 10^{-13}\,\text{m}^2/\text{s}$ at room temperature. Balancing the inertial force with the viscous force yields the momentum relaxation time $\tau_b = \frac{4}{3}\pi a b^2 \rho D_1/k_B T \approx 10^{-7}$ s, where $\rho \simeq 1\,\text{g/cm}^3$ is the mass density of the cell. This time is far too short to be relevant to our experiment.

The above back-of-the-envelope calculations show that the duration of a flick, or the time it takes for the cell body to reorient, $\tau_r \sim 10-20$ ms $\gg \tau_f, \tau_b, \tau_h$, is the longest relaxation time for the cell body reacting to CW→CCW motor reversals. Although translational motions of the cell body can still be resolved for time less than $\tau_r$ by high-speed video imaging [14], this time scale is relevant for our experiment since at the normal video speed, cell reorientation is a major signature of CW→CCW transitions seen under the microscope (see Fig. 6 (C, D) and Ref. [24]).

A swimming interval, $\Delta_f$ or $\Delta_b$, consists of two motor reversals, and therefore the minimal interval length must be $\Delta_{min} \simeq \tau_r$ or 20 ms, which is comparable to our video resolution $\pm 16.7$ ms. We note that the measured $P(\Delta_f)$ and $P(\Delta_b)$ are peaked at $\sim 270$ and $\sim 370$ ms, respectively, which are an order of magnitude greater than $\Delta_{min}$, and these PDFs drop to nearly zero in the neighborhood of $\Delta_{min}$ as seen in Fig. 3. Physically, the elastic hook behaves like a low-pass filter that "masks" those short intervals ($\Delta_s < \Delta_{min}$) when the cell body is unable to respond and "let goes" those long intervals ($\Delta_s > \Delta_{min}$) when the cell body is able to respond, where $s \in \{f, b\}$. The fact that the probability of observing small $\Delta_s$ drops precipitously suggests that inhibition of these short intervals is intrinsic to the flagellar motor switch of *V. alginolyticus*.

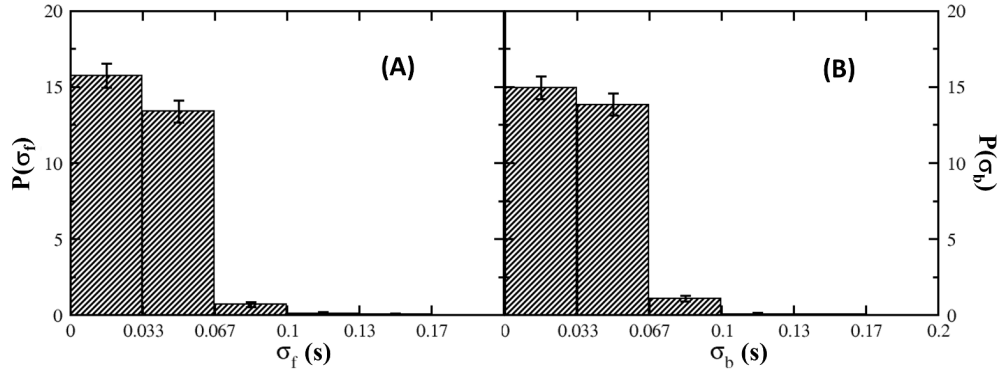

Figure S6: PDFs of $\sigma_f$ and $\sigma_b$. The plots depict the distribution of errors in determining the forward dwell time $\Delta_f$ (A) and the backward dwell time $\Delta_b$ (B). As can be seen, majority of the intervals can be determined with an uncertainty less than 66 ms. The frequency of uncertainties greater than 66 ms is very small.

## Uncertainties in Determining Dwell Times $\Delta_f$ and $\Delta_b$

*V. alginolyticus* swims at an average speed of 55 $\mu$m/s. Therefore during one video frame, $\Delta t = 33$ ms, the average displacement of the cell along its axis is $\sim 1.8\,\mu$m, which is about 7.5 pixels in our setup. This is significantly greater than the displacement by diffusion in the same interval, $\delta_{diff} = \sqrt{2D_1\Delta t} = 0.14\,\mu$m, which is much less than one pixel in our video images. Likewise we can compute the smearing effect due to rotational diffusion of the cell body. The rotational diffusion coefficient $D_2$ is given by [25]:

$$D_2 = \frac{3k_BT}{8\pi\eta a^3}(\ln\frac{2a}{b} - \frac{1}{2}) \simeq 0.19\,\text{rad}^2/\text{s}. \qquad (2)$$

The typical rotation of the cell body due to diffusion over $\Delta t = 33$ ms is $\delta_{rdiff} = \sqrt{2D_2\Delta t} = 0.11$ rad.

Based on the above calculation, we set up an objective criterion for determining the moment of a motor reversal and the associated uncertainty: During a motor reversal, if the displacement of a cell is less than 0.5 $\mu$m ($\sim 3\delta_{diff}$) and the cell body's orientation changes less than 0.33 rad ($\sim 3\delta_{rdiff}$) between two consecutive frames, the motor state during the second frame is considered unknown and the uncertainty in deciding the transition moment increases by $\pm 16.7$ ms. The uncertainties $\sigma_f$ and $\sigma_b$ for each $\Delta_f$ and $\Delta_b$ can thus be obtained, and the result is presented in Fig. S6. The figure shows that majority of motor reversal events occur rapidly with $\sigma_f$ and $\sigma_b$ being less than 66 ms. The frequency of observing large uncertainties drops by more than an order of magnitude for $\sigma_f$, $\sigma_b > 66$ ms. This justifies the resolution limit, the shaded areas in Fig. 3, presented in the main text.

# References

[1] Kawagishi, I., Y. Maekawa, T. Atsumi, M. Homma, and Y. Imae, 1995. Isolation of the polar and lateral flagellum-defective mutants in Vibrio alginolyticus and identification of their flagellar driving energy sources. *J. Bacteriol.* 177:5158–5160.

[2] Tokuda, H., T. Nakamura, and T. Unemoto, 1981. Potassium ion is required for the generation of pH-dependent membrane potential and $\Delta$pH by the marine bacterium Vibrio alginolyticus. *Biochemistry* 20:4198–4203.

[3] Kojima, S., Y. Asai, T. Atsumi, I. Kawagishi, and M. Homma, 1999. Na+-driven flagellar motor resistant to phenamil, an amiloride analog, caused by mutations in putative channel components1. *Journal of Molecular Biology* 285:1537 – 1547.

[4] Magariyama, Y., M. Ichiba, K. Nakata, K. Baba, T. Ohtani, and S. Kudo, 2005. Difference in bacterial motion between forward and backward swimming caused by the wall effect. *Biophys. J.* 88:3648–3658.

[5] Berke, A., L. Turner, H. Berg, and E. Lauga, 2008. Hydrodynamic attraction of swimming microorganisms by surfaces. *Phys. Rev. Lett.* 101:038102.

[6] Fahrner, K., W. Ryu, and H. Berg, 1965. Bacterial flagellar switching under load. *Nature* 423.

[7] Yuan, J., K. Fahrner, and H. Berg, 2009. Switching of the bacterial flagellar motor near zero load. *J. Mol. Biol.* 390:390–400.

[8] Altindal, T., S. Chattopadhyay, and X. Wu, 2011. Bacterial Chemotaxis in an Optical Trap. *PLoS ONE* 6:e18231.

[9] Chattopadhyay, S., R. Moldovan, C. Yeung, and X. L. Wu, 2006. Swimming efficiency of bacterium Escherichiacoli. *Proceedings of the National Academy of Sciences* 103:13712–13717.

[10] Terashima, H., H. Fukuoka, T. Yakushi, S. Kojima, and M. Homma, 2006. The Vibrio motor proteins, MotX and MotY, are associated with the basal body of Na+-driven flagella and required for stator formation. *Mol Microbiol* 62:1170–1180.

[11] Block, S. M., D. F. Blair, and H. C. Berg, 1989. Compliance of bacterial flagella measured with optical tweezers. *Nature* 338:514–518. 10.1038/338514a0.

[12] Sen, A., R. K. Nandy, and A. N. Ghosh, 2004. Elasticity of flagellar hooks. *Journal of Electron Microscopy* 53:305–309.

[13] Bai, F., R. Branch, D. Nicolau, T. Pilizota, B. Steel, P. Maini, and R. Berry, 2010. Conformational spread as a mechanism for cooperativity in the bacterial flagellar switch. *Science* 327:685–689.

[14] Son, K., J. S. Guasto, and R. Stocker, 2013. Bacteria can exploit a flagellar buckling instability to change direction. *Nat Phys* 9:494–498.

[15] Nishitoba, M., N. Imai, Y. Magariyama, and S. Kudo, 2000. Observation of bacterial flagellar deformation with laser dark-field microscope. *47th Spring Meeting of Japan Soc. Appl. Phys. and Related Soc. (in Japanese)* .

[16] Takano, Y., K. Yoshida, S. Kudo, M. Nishitoba, and Y. Magariyama, 2003. Analysis of small deformation of helical flagellum of swimming Vibrio alginolyticus. *JSME Int. J. Ser. C.* 46:1241–1247.

[17] Magariyama, Y., S. Sugiyama, K. Muramoto, I. Kawagishi, Y. Imae, and S. Kudo, 1995. Simultaneous measurement of bacterial flagellar rotation rate and swimming speed. *Biophys J* 69:2154–2162.

[18] Lighthill, J., 1976. Flagellar Hydrodynamics. *SIAM Review* 18:161–230.

[19] Vogel, R., and H. Stark, 2010. Force-extension curves of bacterial flagella. *The European Physical Journal E* 33:259–271.

[20] Hirano, T., S. Yamaguchi, K. Oosawa, and S. Aizawa, 1994. Roles of FliK and FlhB in determination of flagellar hook length in Salmonella typhimurium. *Journal of Bacteriology* 176:5439–5449.

[21] Shibata, S., N. Takahashi, F. F. V. Chevance, J. E. Karlinsey, K. T. Hughes, and S.-I. Aizawa, 2007. FliK regulates flagellar hook length as an internal ruler. *Molecular Microbiology* 64:1404–1415.

[22] Sowa, Y., H. Hotta, M. Homma, and A. Ishijima, 2003. Torque-speed relationship of the Na+-driven flagellar motor of Vibrio alginolyticus. *J Mol Biol* 327:1043–1051.

[23] Chattopadhyay, S., and X. L. Wu, 2009. The effect of long-range hydrodynamic interaction on the swimming of a single bacterium. *Biophysical Journal* 96:2023–2028.

[24] Xie, L., T. Altindal, S. Chattopadhyay, and X. Wu, 2011. Bacterial flagellum as a propeller and as a rudder for efficient chemotaxis. *Proc Natl Acad Sci USA* 108:2246–2251.

[25] Berg, H., 1993. Random Walks in Biology. Princeton University Press, New Jersey.